\documentclass[preprint]{aastex}
\usepackage{epsfig}

\shorttitle{FUSE Survey of CVs} 
\shortauthors{Godon et al.}
\begin{document}
\bibliographystyle{apj}

\title{A Far Ultraviolet Spectroscopic Explorer Survey of 
High Declination Dwarf Novae  
\altaffilmark{1}}

\author{Patrick Godon\altaffilmark{2}, Edward M. Sion} 
\affil{Department of Astronomy and Astrophysics
Villanova University,
Villanova, PA 19085,
patrick.godon@villanova.edu; edward.sion@villanova.edu}

\author{Paul E. Barrett} 
\affil{United States Naval Observatory, 
Washington, DC 20392
barrett.paul@usno.navy.mil} 

\author{Paula Szkody}
\affil{Department of Astronomy,
University of Washington,
Seattle, WA 98195,
szkody@astro.washington.edu} 

\altaffiltext{1}
{Based on observations made with the 
NASA-CNES-CSA Far Ultraviolet Spectroscopic
Explorer. {\it{FUSE}} is operated for NASA by the Johns Hopkins University under
NASA contract NAS5-32985} 
\altaffiltext{2}
{Visiting at the Space Telescope Science Institute, Baltimore, MD 21218,
godon@stsci.edu}

\begin{abstract}

We present a spectral analysis of the Far Ultraviolet Spectroscopic
Explorer ({\it{FUSE}}) spectra of eight high-declination 
dwarf novae (DNs) obtained from a Cycle 7 {\it FUSE} survey. 
These DN systems have not been previously studied in the UV and
little is known about their white dwarfs (WDs) or accretion disks. 
We carry out the spectral analysis of the {\it FUSE} data using synthetic
spectra generated with the codes TLUSTY and SYNSPEC.
For two faint objects (AQ Men, V433 Ara) we can only assess 
a lower limit for the WD temperature or mass accretion rate.
NSV 10934 was caught in a quiescent state and its spectrum is consistent
with a low mass accretion rate disk. 
For 5 objects (HP Nor, DT Aps, AM Cas, FO Per and ES Dra) 
we obtain WD temperatures between 34,000K and 40,000K and/or  
mass accretion rates consistent with intermediate to outburst states. 
These temperatures reflect the heating of the WD due to on-going
accretion and 
are similar to the temperatures of other DNs observed on the rise to,  
and in decline from outburst.
The WD Temperatures we obtain should therefore be considered as upper limits,  
and it is likely that during quiescence  
AM Cas, FO Per and ES Dra are near the average WD $T_{eff}$
for catalcysmic variables 
above the period gap ($\sim$30,000K), similar to U Gem, 
SS Aur and RX And.  

\end{abstract}

\keywords{Accretion, Accretion disks, Stars: white Dwarfs, Stars: dwarf novae }  

\section{Introduction} 

\subsection{Cataclysmic Variables} 

Cataclysmic variables (CVs) are 
semi-detached binary systems, in which a white dwarf 
star (WD) accretes matter from 
a main-sequence (MS, or post-MS; the {\it secondary}) star.
Accretion takes place via a disk for non-magnetic WDs, and 
via a column or curtain for magnetic WDs   
(for a review on CVs see \citet{lad93,war95}). 
The luminosity of a CV can vary significantly, 
and systems can be found in a state of 
low luminosity ({\it quiescence}),  
high luminosity ({\it outburst}), or 
even in a rather {\it intermediate} state when the system is
{\it active} without reaching its peak luminosity 
\citep{rob76,lad93,war95}.  
CVs are divided in sub-classes according to the duration, occurrence
and amplitude of their outbursts. 

The two main types of CVs are  
dwarf nova (DN) and nova-like (NL) systems. 
Dwarf novas are weakly- or non-magnetic disk systems,
divided into U Gem systems, 
SU UMa systems, and Z Cam systems. 
The U Gem systems are the typical DNs, i.e. those systems exhibiting
normal DN outbursts; 
the SU UMas exhibit both normal DN outbursts, and
superoutbursts, which are both longer in duration and higher in 
luminosity than normal DN outbursts; 
and the Z Cam systems have standstills where they remain in a semi-outburst
state for a long time.
Nova-likes form a less homogeneous class,  
e.g. non-magnetic disk systems found mostly in high state
(known anti-DN or VY Scl), systems that exhibit no known low
states (UX UMa systems),  
non-disk magnetic systems (AM Her stars or Polars), 
magnetic systems with truncated disk (Intermediate Polars), etc.  
See e.g. \citet{rit03} for the classification of CVs .

The binary orbital period in CVs 
ranges from a fraction of an hour to about a day; 
however, there is a gap in the orbital period
between 2 and 3 hours where almost no CV systems are found 
(hereafter the ``period gap'' \citep{why80}). 
U Gem and Z Cam DN systems are found above the period gap, 
while the SU UMa DN systems are found below the period gap.  

\subsection{The Standard Model} 

In the commonly accepted (standard) theory of evolution of CVs,
systems
evolve through the gap from long periods to shorter periods, driven by
angular momentum loss. Above the gap ($P>3$hrs) magnetic braking
\citep{web67,ver81}
is thought to be the dominant mechanism for angular momentum loss 
due to wind loss of the tidally locked secondary. 
At about $P\approx3$hrs, 
magnetic braking stops when
the secondary becomes fully convective, and its radius becomes smaller
than the Roche lobe radius: the secondary detaches from the Roche
lobe. At this point, gravitational radiation \citep{fau71,pac81,rap82} 
is believed to take over
as the main mechanism for angular momentum loss. The binary separation 
and period keep on decreasing until $P\approx 2$hrs,
at which time the Roche lobe radius becomes again comparable to the
radius of the secondary, and the binary becomes again a contact binary.   
The disrupted (or {\it interrupted}) 
magnetic braking became the standard scenario of CV
evolution explaining the presence of the period gap around 3hrs 
\citep{pac83,rap83,spr83,ham88,kin88,kol93,how01}.   

The first comparison of CV SD temperatures with actual
evolutionary tracks (the evolutionary tracks of \citet{mcd89})
was carried out by \citet{sio91}, 
who compared the CV WD $T_{eff}$ values as a function of 
orbital period with the cooling curves of non-accreting WDs
and with the theoretical prediction of $\dot{M}$ (hence $T_{eff}$)
versus $P$ of \citet{mcd89}.  
\citet{sio91} found some evidence, 
though far from conclusive, that CV systems evolve across the 
period gap.
However, the standard model is based on an ad-hoc assumption about the
mechanism of orbital angular momentum loss, and many of its
predictions are in disagreement with the observations \citep{tow09}.
For example, if CVs are evolving from longer period to shorter period across
the gap, then short period DNs are older and with a longer history of 
angular momentum transfer via disk accretion, and their WDs should 
rotate faster than DN WDs above the gap. 
However, so far, there is no observational evidence pointing to this.  
Modifications have been proposed to the standard model 
\citep{bas87,ham90,mcc98,sche98,kol99,cle98,
spr01,taa01,kol01,kin02,taa03,iva04}.  
One of the alternatives has even been 
the suggestion that CVs above and below the gap form two
distinct populations not related to each other
\citep{egg76,web81,and03b}. 

In order to put more constraints 
on the theories of the evolution of CVs, more observations
are needed \citep{sio91,gan05}.   
The standard model gives a prediction of
$\dot{M}$  above the period gap (as a function of the binary period)
and as a consequence $\dot{M}$ is the most important parameter 
to test the theory. 
Unfortunately, to derive $\dot{M}$ from observations
proves to be difficult, as it itself relies on many unknowns (e.g. 
the distance). 
For example, optical observations seem to indicate that 
$\dot{M}$ is correlated with the period \citep{pat84},
while FUV spectroscopy (albeit of NLs alone) 
does not show any correlation at all \citep{pue07}. 

\subsection{The White Dwarf Temperature: the Case for Dwarf Novae} 

Recently,  \citet{tow03,tow04} have shown that
the quiescent effective temperatures $T_{eff}$ of the 
WD in CVs  approaches an 
equilibrium value when the WD is subject to accretion at an  
average accretion rate 
$\dot{M}$  which is constant on time scales similar to the WD core 
thermal time ($\sim$100Myrs).  
This equilibrium value is a function of the WD mass and
mass accretion rate only.  
Therefore, knowledge of the quiescent surface temperature 
of the WD and its mass
can be used to obtain an estimate of the average mass accretion rate  
(it seems counter-intuitive that the effective temperature of the
WD would not depends on its age; 
the limitations and the applications of the relation 
between $<\dot{M}>$ and $T_{eff}$ are discussed in \citet{tow09}).   
As a consequence, CV WD temperature has become the most important 
observational parameter, together with the binary period and
WD mass, to test the theories of the evolution of CVs. 

For DNs below the gap, the distribution of WD temperatures 
obtained from FUV spectroscopic analysis  
seems to be centered around $T_{wd} \sim$15,000K, 
while for systems above the gap the WD temperatures are higher,
somewhere between $\sim 30,000$K and up to $\sim 50,000$K 
\citep{sio08}.  Above the gap at periods of about 200-300min, 
the WDs in DNs are about 10,000K cooler than the WDs in VY Scl systems  
\citep{god08b}.
On the other hand, polars (devoid of disks)  
have the coolest WDs of all, 
with $T_{wd}<20,000K$ above the gap and $T<15,000$K below
the gap \citep{ara05} (see \citet{tow09} for a review WD temperatures
in CVs).

Currently, the number of white dwarf temperatures known  are
as follows.  
There are 10 NL AM Her systems below the gap
and 3 above the gap; 5 NL VY Scl system above the gap;  
there are 19 DN SU UMa systems below the gap;
12 DN U Gem systems above the gap;
and only 2 DN Z Cam systems above the gap. 
There is a critical shortage 
in knowledge of the WD effective temperature $T_{eff}$ 
for DN Z Cam, NL AM Her and NL VY Scl all 
above the period gap.  
Thus, detailed comparisons of accreting WDs above and 
below the gap cannot be made. 

However, during quiescence DNs, unlike other CVs, have the advantage 
of exposing their WD in the FUV for long periods of time. 
With a temperature range $\sim 12,000-50,000$K, the WD
peaks in the FUV, and this makes {\it FUSE} the best instrument 
to observe the WD of quiescent DNs.  
Furthermore, DNs offer a fairly reliable estimate of
their distances via the absolute magnitude at maximum versus orbital
period relation found by \citet{war87} and \citet{har04}. 
As a consequence,  DNs are ideal candidates to observe with {\it FUSE}, 
as one can more easily derive the WD temperature. 

More precisely, 
by fitting the observed quiescent spectrum with synthetic spectra, 
one can, in theory, derive the WD parameters such as the 
temperature, gravity, rotational velocity, and chemical
abundances (mainly for C, N, S, Si). 
The (instantaneous) mass accretion rate of many systems 
can be deduced too at given epochs
of outburst or quiescence using spectral fitting
techniques of FUV spectra. 
In practice, however, to obtain robust results in 
the spectral analysis and depending on the quality of the
observed spectrum, one needs to know the distance $d$ to the system, 
the WD mass $M_{wd}$, and possibly also the inclination of
the system (see sections 3 and 4).   

For these reasons, 
we present here a spectroscopic analysis of a 
sample of dwarf novae from our  Cycle 7 {\it FUSE} survey.
The present analysis includes 9 dwarf nova systems:
6 systems above the gap (including 4 Z Cam),   
1 below the gap, and 2 with unknown period.   
The observations are described in the next section, including details 
of the processing of the spectra and a general description of the lines
seen in {\it FUSE} spectra of DNs. In section 3 we present our 
synthetic spectral modeling that we use to analyze the spectra. 
In section 4 we present and discuss the results of the spectral
analysis for each system separately. And in the last section we
give a short summary of our findings.

\section{Observations} 

\subsection{The {\it FUSE} Spectra} 

The FUSE targets    are listed in Table 1 with their system 
parameters as follows: 
column 
(1) Name, 
(2) CV subtype,
(3) orbital period in days, 
(4) orbital inclination in degrees, 
(5) apparent magnitude in outburst, 
(6) apparent magnitude in quiescence, 
and (7) priority given as a {\it FUSE} target (1=high, 2=low).

The observation log is presented in Table 2. 
The following systems were also listed as targets in our original 
proposal, but they were not observed with {\it FUSE}: 
AD Men, V342 Cen, V1040 Cen, and TZ Per.   
From the {\it{AAVSO}} (American Association of Variable Stars Observers)
data \citep{pri06}, we were able to assess the state of each target at the
time of the FUSE observations.  

All the {\it{FUSE}} spectra were obtained 
through the 30"x30" LWRS Large Square Aperture in TIME TAG mode. 
The data were processed with the latest and final version of 
CalFUSE (v3.2.0; for more details see \citet{dix07}).
The {\it{FUSE}} data comes in the form of eight spectral segments 
(SiC1a, SiC1b, SiC2a, SiC2b, LiF1a, LiF1b, LiF2a, and LiF2b) which
are combined together to give the final {\it FUSE} spectrum.   
To process {\it{FUSE}} data, we follow the same procedure we used  
previously for the analysis of other systems (such as WW Ceti, \citet{god06a});
consequently we give here only a short account of this procedure.  
The spectral regions covered by the
spectral channels overlap, and these overlap regions are then used to
renormalize the spectra in the SiC1, LiF2, and SiC2 channels to the flux in
the LiF1 channel. We then produced a final spectrum that covers the
full {\it{FUSE}} wavelength range $905-1187$ \AA. The low sensitivity 
portions of each channel were discarded.
In most channels there exists a narrow dark stripe of decreased flux
in the spectra running in the dispersion direction, 
known as the ``worm'', which can attenuate as much as
50\% of the incident light in the affected portions of the
spectrum; - this is due to shadows thrown by the wires on the grid
above the detector. In the present observations, the worm affected
mainly the 1bLiF channel.  
Because of the temporal changes in the strength and position of the 
``worm'', CALFUSE cannot correct target fluxes for its presence. 
Therefore, we carried out a visual inspection of the {\it{FUSE}} channels to
locate the worm and we {\it{manually}} discarded the 
portion of the spectrum affected by the worm.
We combined the individual exposures and channels to create a
time-averaged spectrum weighting
the flux by the exposure time and sensitivity of the
input exposure and channel of origin. 

For all the targets we had to remove 
the edges of the SiC channels and the worm ($\sim$1140-1180\AA ) in the 
1bLiF channel. In some cases we had to remove
entire channels due to their very poor S/N.   
Specifically, we proceeded as follows: \\  
- AQ Men, we kept only the 2aLif and 2bLif channels;  \\ 
- HP Nor, we removed all the SiC channels while keeping the LiF channels;  \\ 
- DT Aps, we removed large portions (at the edges) of the SiC channels; \\ 
- AM Cas, we discarded the 2bLiF, 2bSiC, 2aLiF and 2aSic channels and
only used 1aSiC, 1bSiC, 1aLiF, and 1bLiF;  \\ 
- NSV 10934, we removed 1aSic, 2bSiC, 2bLiF;  \\ 
- FO Per, we kept all the channels;  \\ 
- ES Dra, we kept all the channels; and  \\ 
- V433 Ara, all the SiC channels were discarded 
and only the LiF channels were kept.   \\

\subsection{The {\it{FUSE}} Lines} 

The main emitting components contributing to the {\it{FUSE}} spectra of DNs 
are the accretion disk and the WD: the disk dominates
during outburst, while the WD dominates during quiescence. The main feature
of the  spectra is the broad Ly$\beta$ absorption feature, which can
easily be used to assess the gravity and temperature of the emitting
gas. 
At higher temperatures,  
as the continuum rises in the shorter wavelengths,
the higher orders of the Lyman series also become visible; however,  
they become narrower. 

Additional
broad absorption lines of metals (C, S, Si, ..) are detected 
and help determine the chemical abundances and projected rotational
velocity of the emitting gas.   
In the present spectra, the main absorption features observed 
are: 
C\,{\sc ii} (1010 \AA ),  
C\,{\sc iii} (1175 \AA ),   
Si\,{\sc iii} ($\approx$1108-1114 \AA\ and  
$\approx$1140-1144 \AA ), 
Si\,{\sc iv} ($\approx$1067 \AA\ and $\approx$1120-1130 \AA\ ), 
S\,{\sc iv} (1073 \AA ),  
and 
N\,{\sc ii} (1085 \AA\ when not contaminated by air glow).   

On top of the spectrum, broad emission lines are also found in two  of
the systems, NSV 10934 and AQ Men,  mainly the  O\,{\sc vi} doublet 
and C\,{\sc iii} (977\AA\ and 1175 \AA ). 
The broad emission lines in NSV 10934 and AQ Men originate  
from the hot accretion region on the white dwarf, and 
possibly also from the inner disk itself, as expected for high-inclination
DNs in quiescence such as BZ UMa \citep{gan03} or EK TrA \citep{god08a}.    
AQ Men is eclipsing and the presence of broad emission lines in NSV
10934 might be an indication that it is a high inclination system.

All our {\it{FUSE}} spectra 
show some ISM molecular hydrogen absorption. The most affected
targets reveal a spectrum ``sliced'' 
at almost equal intervals ($\sim$12\AA );  
starting at wavelengths around 1110\AA\ and continuing
towards shorter wavelengths all the way down to the hydrogen cut-off around
915\AA\ . In the affected {\it{FUSE}} spectra, we identified the 
most prominent molecular hydrogen absorption lines by their band
(Werner or Lyman), upper vibrational level (1-16), and rotational transition
(R, P, or Q) with lower rotational state (J=1,2,3).

In addition,    the targets that are weak 
{\it{FUSE}} sources exhibit sharp emission lines from air 
glow (geo- and helio-coronal in origin; some of which is due to sunlight
reflected inside the telescope),  such as 
H\,{\sc i} series, S\,{\sc vi} (934, 944),  O\,{\sc vi} doublet, 
C\,{\sc iii} (977), and He\,{\sc i} (1168).  

For each system, we mark on the figures the lines that are detected
as well as lines that are commonly seen for comparison. 
The lines that are commonly seen in the {\it FUSE} spectra of 
quiescent DN systems are as follows. 
At lower temperatures ($ 15,000$K$<T<25,000$K), the 
observed lines are (e.g. \citet{god08a}) 
C\,{\sc iii} (1175 \AA ),
C\,{\sc ii} (1066 \AA ),
Si\,{\sc iii} ($\approx$1108-1114 \AA\ and
$\approx$1140-1144 \AA ), and
N\,{\sc ii} (1085 \AA\ when not contaminated by air glow).
At higher temperatures (T$>25,000$K),
as the continuum rises in the shorter wavelengths,
the higher orders of the Lyman series are revealed. 
At these temperatures the S\,{\sc iv} (1073 \AA )
absorption line starts to appear, and, as
there is more flux in the shorter wavelengths,
the  C\,{\sc ii} (1010 \AA ) absorption line also becomes visible 
(e.g. \citet{urb06,lon99}).
At still higher temperature (T$>50,000$K), the
C\,{\sc ii} and  Si\,{\sc iii} lines disappear, and the spectrum becomes
dominated by high order ionization lines such as
N\,{\sc iv} ($\approx$923 \AA ),  S\,{\sc vi} (933.5 \& 944.5 \AA ),
S\,{\sc iv} (1063 \& 1073 \AA ),  Si\,{\sc iv} (1066.6 \AA ),
and O\,{\sc iv} (1067.8 \AA ) (see e.g. \citet{har05,sio07}).
 
Because
of the low quality of the spectra, the observed wavelengths 
cannot be determined accurately and, therefore, we do not list 
the wavelengths of the lines in a table. The lines are discussed
for each object separately. 

\section{Spectral Modeling} 
 
\subsection{The Synthetic Stellar Spectral Codes}

Before we carry out the actual fit for each individual observed spectrum,
we create a grid of model spectra for
high-gravity stellar atmospheres using codes 
TLUSTY and SYNSPEC\footnote{
http://nova.astro.umd.edu; TLUSTY version 200, SYNSPEC version 48} 
\citep{hub88,hub95}. 
Atmospheric structure is computed (using TLUSTY) assuming a H-He LTE 
atmosphere; the other species are added in the spectrum synthesis
stage using SYNSPEC. 
For hot models (say $T>50,000$K) we switch the approximate NLTE treatment 
option in SYNSPEC (this allows 
us to consider and approximate NLTE treatment
even for LTE models generated by TLUSTY). 
We generate photospheric models with effective
temperatures ranging from 12,000K to 75,000K in increments of 
about 10 percent (e.g. 1,000K for T$\approx$15,000K and 5,000K for 
T$\approx$70,000K).
As the WD mass in these DN systems is  not known,
we choose values of $Log(g)$ ranging between 7.5 and 9.5. 
We also vary the
stellar rotational velocity $V_{rot} sin(i)$ from $100$km$~$s$^{-1}$
to $1000$km$~$s$^{-1}$ in steps of $100$km$~$s$^{-1}$. 
As a default, solar abundances are assumed for all the models in the grid.  
We have, however, the option to vary the chemical abundances of individual 
elements in the models in the grid.  
For any WD 
mass, there is a corresponding radius, or equivalently, one single value of
$Log(g)$ (e.g. see the mass radius relation
from \citet{ham61} or see \citet{woo90,pan00} for different
composition and non-zero temperature WDs).

The same suite of code is also used to generate spectra of
accretion disks \citep{wad98} based on the standard model of 
\citet{sha73}. In the present work we use disk
spectra from the grid of spectra generated by \citet{wad98}
as well as disk spectra that we generate. A detailed description
of the procedure to generate such disk spectra is given 
in \citet{wad98}. 

\subsection{Synthetic Spectral Model Fitting}

Before carrying out a synthetic spectral fit of the spectra,
we masked portions of the spectra with strong emission lines,
strong ISM molecular absorption lines, detector noise and air glow.
These regions of the spectra are somewhat different for each object and 
are not included in the fitting. The regions excluded from the fit
are in blue in the figures.
The excluded ISM quasi-molecular absorption lines are 
marked with vertical labels in the figures.

After having generated grids of models for each target, 
we use FIT \citep{numrec}, a $\chi^2$ minimization routine,
to compute the reduced $\chi^{2}_{\nu}$ 
($\chi^2$ per number of degrees of freedom) 
and scale factor values for each model fit.  
While we use a $\chi^2$ minimization technique, we do not 
blindly select the least $\chi^2$ models, but we examine the models 
that best fit some of the features such as absorption
lines (see the fit to the {\it{FUSE}} spectrum alone) 
and, when possible, the slope of the wings of the broad Lyman
absorption features. When possible, we 
also select the models that are in agreement
with the distance of the system (or its best estimate).  

The flux level at 1000\AA\ (between Ly$\delta$ and Ly$\gamma$) 
is close to zero for temperatures below 18,000K, at 30,000K it is about
50\% of the continuum level at 1100\AA\ and it reaches 100\% for T$>$45,000K.  
At higher temperature (T$>$50,000K) the spectrum becomes pretty flat 
and there is not much difference in the shape of the spectrum 
between (say) a 50,000K and a 80,000K model.  
When fitting the shape of the spectrum in such a manner, 
an accuracy of about 500-1,000K is obtained, due to the S/N.
In theory, a fine tuning of the temperature (say to an accuracy of about
$\pm$50K) can be carried
out by fitting the flux levels such that the distance to the system
(if known) is matched. However, the fitting to the
distance depends strongly on the radius (and therefore the mass)
of the WD. In all the systems presented here the mass
of the WD and the distance are 
unknown and therefore we are unable to assess the
temperature accurately.  
Furthermore, since the Ly$\beta$ profile  depends on both
the temperature and gravity of the WD, the accuracy of the
solution is further decreased as there is a degeneracy in the solution,
namely the solution spreads over a region of the $Log(g)$ and $T$
parameter space. And last, reddening values are 
rarely known, therefore increasing even more      
the inaccuracy of assessing the temperature by scaling the synthetic flux
to the observed flux. As a consequence, for each target we present 
more than one model fit.  

For DT Aps and ES Dra we vary the abundances; however,   
because of the low S/N, for all the other spectra 
we fit solar abundance models and do not change the 
abundance of the species.

The WD rotation ($V_{rot} sin(i)$) rate is determined by fitting 
the observed {\it FUSE}  spectrum with WD models with increasing
values of the rotational velocity.  
We did not carry out separate fits to individual lines but
rather tried to fit the lines and continuum in the same fit.

For each spectrum, when possible, we try to fit 
a single WD model, a single disk model, and a composite WD+disk model,
assuming different reddening values. For all the systems presented
here the WD mass is unknown and the distance is estimated using the
maximum magnitude/period relation \citep{war87,har04}, or (when the
period is unknown) the method described by \citet{kni06,kni07} 
(see Table 3). 
  
For the single WD model, we vary the temperature while first keeping
the WD mass constant, starting at about $0.4M_{\odot}$. 
Once the lowest $\chi^2$ has been found for a given mass, we 
vary the projected rotational velocity, and possibly also the abundances, 
to further lower the $\chi^2$ and obtain a best fit.  
Once the best fit has been found for that mass, 
we assume a slightly larger
mass and again vary the temperature until the lowest $\chi^2$ is found. 
We follow this procedure iteratively 
until we reach a mass of about $1.2M_{\odot}$. 
The next step is to find, from all these lowest $\chi^2$ models, which
one agrees best with the distance estimate, or which one has the
lowest $\chi^2$ of all (if the constraint on the distance cannot be
used). For the single disk model, we carry out a similar procedure by 
varying the mass accretion rate and inclination assuming discrete values
of the WD mass, and then chose the least $\chi^2$ model agreeing best
with the distance. And last, we use the same procedure 
for the WD+disk composite modeling to find the best fit model.

\section{Results and Discussion}

Four systems were caught in deep quiescence, and their {\it FUSE}
spectra have such a low S/N that they are unusable. This is due not only
to  the intrinsic low brightness of the targets, but also to the loss of data
in some {\it FUSE} channels (mainly SiC) and to the loss of good
exposure time (due e.g. to jitters). 
These four systems are VW Tucanae/HV 6327, IK Normae, V663 Arae/S 5893, 
and V499 Arae/S 6115. 
The remaining eight systems that were successfully observed 
are presented here.

\subsection{Southern Hemisphere Systems} 

\subsubsection{HP Normae/HV 8865} 
HP Nor is a DN Z Cam system \citep{vog82} which exhibits 
coherent oscillations 
during outbursts with a period $\sim 18.6$s \citep{pre06}.  
Its magnitude varies between V=12.6 at maximum and V=16.4 
at minimum \citep{bru94}.  
The distance to HP is not known, and since its period is
also unknown, we have no way to assess its distance.
HP Nor was observed with {\it FUSE} on April 13 (JD 2454203.7).   
Data from the AAVSO and AVSON imply that the system was 
in a low state of brightness $V\sim 15$, though it did not
reach its lowest magnitude $V=16.4$.   
The continuum flux level of the {\it FUSE} spectrum is roughly consistent
with the continuum flux level in the optical around 3,200\AA\  
obtained by \citet{mun98}, who observed the system in an 
apparent decline from outburst with V=14.18 and a flux level 
of $2\times 10^{-14}$ergs$~$cm$^{-2}$sec$^{-1}$\AA$^{-1}$.    
The {\it FUSE} spectrum of HP Nor is presented in 
Figure 1.
The reddening toward HP Nor is not known, however the 
Galactic reddening in this direction is 0.63. 
Therefore, in the following 
we model HP Nor assuming both E(B-V)=0.0 and E(B-V)=0.2 . 

\paragraph{E(B-V)=0.0} 
For the single WD model, assuming $M_{wd}=0.8M_{\odot}$,  
we find a temperature of 37,000K and a distance of 648pc. 
The projected rotational velocity is $v_{rot} \sin{i} = 1000$km/s 
to fit the C\,{\sc iii} (1175) line profile.
The $\chi^2$ obtained is smaller than one due to the poor S/N, 
namely $\chi^2_{\nu} =0.107$. 
The results are listed in Table 4, where we also present  
a $0.4M_{\odot}$ WD and a $1.2M_{\odot}$ WD models. 
For the single disk models, assuming  
a $0.8M_{\odot}$ WD mass, we find                 
$\dot{M}=10^{-9.0}M_{\odot}$/yr, $i<41^{\circ}$, d=1603pc and
$\chi^2_{\nu}=0.104$. 
Again, in Table 4 we list the results assuming different values
for $M_{wd}$.
There is basically no difference in the quality of the fit between
the disk model and the single WD model.  
A WD+disk composite model does not improve
the fit. Because the WD mass and the distance are unknown, we 
are left with solutions spanning a large area in the parameter space,
especially for the composite WD+disk models. We 
do not list the results for the composite models as there is clearly 
a degeneracy in the solution. 

\paragraph{E(B-V)=0.2} 
Dereddening the spectrum for E(B-V)=0.2 increases the temperature of
the single WD model by 3,000K and reduces the distance by a factor of 2.
The $\chi^2$ obtained is slightly lower $\chi^2_{\nu} \approx 0.096$. 
For the single disk models, the mass accretion rate obtained is larger  
by a factor of up to $\sim30$, while the distance is reduced by 
about half (all the models are listed in Table 4).  
Again, there is not much difference between
the single disk model and the single WD model.  
The WD+disk composite models again do not bring any improvement 
in the fit, rather the opposite. 

It is most likely that the spectrum is due solely 
to the heated WD as the system was observed in a relatively low state. 
The best disk models reach a mass accretion rate that is too large
for such a low state. 
Our preferred model is the single WD model for E(B-V)=0.2, assuming
an intermediate mass of $0.8M_{\odot}$ with $T=43,000$K. 
This model is shown in Figure 2. 

\subsubsection{AQ Mensae/EC 05114-7955} 

AQ Men (also known as EC 05114-7955 \citep{che01}) 
has a magnitude around 14-15, but no 
outburst has been detected so far. 
Its period is 0.141466d \citep{pat02}.
Its optical spectrum suggests it is a DN, however, the possibility
that  it is a NL cannot be ruled out \citep{che01}. 
\citet{che01} classified it as an {\it unusual/peculiar} CV, 
that shows large-amplitude
flickering and a low-excitation spectrum, where the 
C\,{\sc iii}/N\,{\sc iii}$\lambda$4650\AA\ blend is also seen in emission,
and suggested that 
the double peak in the line profiles might indicate that it is a highly
inclined system. \citet{pat02} showed that it is actually an eclipsing
system with superhumps and a precession period of 3.77$\pm$0.05d.   

A {\it FUSE} spectrum of AQ Men was obtained on 22 November 2006,
at which time we only know it was at a magnitude between 14 and 15 
(AAVSO data). For the modeling,  
we do not know {\it a priori} whether the disk or the WD dominates the
FUV spectrum.
In spite of its very low continuum flux level 
(at most $\sim 1\times 10^{-14}$ergs$~$cm$^{-2}$sec$^{-1}$\AA$^{-1}$;
see Figure 3) the spectrum 
(binned at 0.2\AA ) exhibits features
that are easily identified  such as the molecular hydrogen absorption 
lines, some Si\,{\sc iii} \& Si\,{\sc iv} and C\,{\sc iii} absorption
features, as well as broad emission from the O\,{\sc vi} doublet
and the C\,{\sc iii} (1175) multiplet. The emission is consistent with
DNs observed in quiescence, and the large (rotational) broadening 
of the C\,{\sc iii} emission ($>$6\AA, or $v>1,000$km/s) agrees   
with the fact that it is highly inclined. However, broad emission 
lines are often observed in nova-likes systems too. 
In the present case the {\it FUSE} spectrum of AQ Men is more 
similar to that of a quiescent DN such as EK TrA \citep{god08a}, as the
emission lines are not as strong as those exhibited by nova-like
systems (such as AE Aqr; see the {\it MAST} archive). 

The distance to AQ Men (using the relations of 
\citep{war87,har04};see Table 3) is possibly $d\approx 710$pc. 
AQ Men has no known reddening but the Galactic reddening
in its direction is $E(B-V)=0.18$. 
We model its {\it FUSE} spectrum first assuming no reddening,
and then assuming $E(B-V)=0.10$. Since it is apparently an eclipsing
system \citep{pat02} we set the inclination angle to $i=81^{\circ}$.  
The spectrum of AQ Men is so poor, that we only try to fit the flux level
in order to obtain the distance to the system irrelevant of the value
of the $\chi^2$. Since we only fit the flux level, we do not show any
figure of the results.  

\paragraph{Single Disk Model.}
For the single disk model fit, we find that, in order to fit the distance, 
the mass accretion rate has to be $\dot{M}=10^{-9.5}M_{\odot}$yr$^{-1}$  
assuming a WD mass of $1.2M_{\odot}$; and it  
increases to $\dot{M}=10^{-8.75}M_{\odot}$yr$^{-1}$ 
for $M_{wd}=0.35M_{\odot}$ 
Next, we carried the same model fits but we deredden the spectrum
assuming $E(B-V)=0.1$. We obtain larger mass accretion rates
as listed in Table 4.  
From these single disk models, we note that if the emission is due to a disk,
then its accretion rate is consistent with the outburst state of a DN, where
the minimum value obtained 
($\dot{M}=10^{-9.5}M_{\odot}$yr$^{-1}$) is for the larger mass
($1.2 M_{\odot}$) assuming no reddening. This is the lower limit of $\dot{M}$
for all reasonable values of $M_{wd}$ ($\le1.2M_{\odot}$) and $E(B-V)\ge0.0$.  

\paragraph{Single WD Model.}
Assuming no reddening and for a mass $M=0.4 M_{wd}$, we fit the flux with 
$T=24,000$K. Increasing the mass decreases
the radius and therefore the flux is fit with a larger temperature 
reaching 36,000K for a $1.1M_{\odot}$ WD. 
Dereddening the spectrum assuming $E(B-V)=0.10$, we obtain
higher temperatures, with 
$T=60,000$K for $M=1.1M_{\odot}$.
From the single WD modeling we obtain that if the spectrum is due to the
WD alone, then its temperature must be $T\ge24,000$K and it increases
with increasing mass and reddening. This is therefore the lower limit 
for the WD temperature of AQ Men, assuming it spectrum is solely from
the WD. 
Since there is no additional constraints on the parameters of the system, 
we are unable to carry out a multi-component (WD+Disk) model fit as 
there is degeneracy. 
However, we note that, because of its inclination 
\citep{pat02}, the disk masks the WD, except if it is 
in a quiescent-like state.  

For AQ Men we can summarize these results as follows. 
If the system is a DN, then it is probably in a lower 
state, where the WD dominates the FUV and has 
a temperature of at least 24,000K.
The disk models do not agree with a quiescent mass accretion rate
for DN.
If the system is a NL, then the disk is in outburst and dominates the FUV
with a mass accretion rate $\dot{M} \le 3 \times 10^{-10}M_{\odot}$yr$^{-1}$, 
barely larger than $\dot{M}_{critic}$ but consistent with the theory.   

\subsubsection{V433 Arae/S 6006} 

V433 Ara is an eclipsing Z Cam DN system, with a period of 
4.698h \citep{wou05}. From the AAVSO data, we find that  
the system was in quiescence at the time of the {\it FUSE} observations,
on 18 Feb 2007.
However, its exact magnitude during the {\it FUSE} observations
is not known.  
The {\it FUSE} spectrum of V433 Ara (Figure 4) is the poorest spectrum of 
all the observed spectra we model, 
it consists only of the LiF channels, and consequently
we are not able to model it properly. 
We model only the flux level to obtain
some constraints on the temperature and mass of the WD and  
possibly on the mass accretion rate. 
The galactic reddening in the direction of V433 Ara is about 0.2
but the actual reddening of V433 Ara is not known. We assume 
here E(B-V)=0.0 and 0.1. Also from the period-luminosity
relation we find a distance of $\sim$1200pc.
Since the system was not observed in outburst      
the spectrum could be either that of quiescence ($\sim$18mag) with
the WD dominating the spectrum, or 
standstill ($\sim$16.5mag) at an intermediate level 
\citep{wou05} where the inflated disk dominates the spectrum (because of the
high inclination). 
We model the spectrum with single disk 
and single WD models, fixing d=1200pc and 
$i\sim 80^{\circ}$.  Since the spectrum is of such poor quality
(see Figure 4) we do not show the model, as we only
fit the {\it level of the continuum} (i.e. there are no 
features to match).   

\paragraph{Single Disk Models.} 
The main results for the disk model are presented in Table 4. 
For a larger WD mass the mass accretion needed to fit the spectrum
decreases. 
If the system was observed with {\it FUSE} in a standstill, 
then the mass accretion rate should rather be on the lower side
in agreement with a large WD mass and little reddening. 
However, because of the large distance, 
the reddening might not be negligible and we therefore suspect that V433 Ara
could have been observed in quiescence and that its FUV flux might be due to
a WD rather than a disk.

\paragraph{Single WD Models.}
The WD models are listed in Table 4. The main result 
is that the temperature of the WD 
increases with the mass of the WD and the reddening. 
From these, it appears that the lower limit
for the WD temperature of V433 Ara is 24,000K. 
Again, if the reddening is not
negligible (E(B-V)=0.1 or larger),  then the WD mass
has an upper limit of about $0.8-1.0 M_{\odot}$, as it is very unlikely
that the WD has a temperature $T\sim 75,000$K (most DNs's WDs 
have $T<50,000$K).    

\subsubsection{DT Apodis/S 5646} 

DT Aps was first identified as a DN by \citet{vog82}, and it 
is classified as a       
SS Cyg sub-type \citep{dow01}. Apart from that little is known 
about this high declination southern dwarf nova.  

DT Aps was observed 
with {\it FUSE} on 28 April 2006 and on 3 May 2006. 
These observation were taken 4 and 9 days (respectively) after an outburst 
recorded by the AAVSO on 24 April 2006.
It is not known how long the
outburst lasted, but it is likely that the first observation
caught the system  while it was still in a relatively active state.  
The second observation was taken after the system had more time
to decline to quiescence, however, 
it was 10 times shorter than the first one 
and it is therefore unusable. 
Even though the first observation lasted basically 70ks, the SiC
channels were very noisy and large portions of the spectral segments  
had to be discarded. The {\it FUSE} spectrum of DT Aps is shown in 
Figure 5.

In the shorter wavelengths (upper panel)  
the spectrum consists mainly of sharp air glow emission lines (noise) with
some possible C\,{\sc iii} ($\lambda 977$) emission; at longer 
wavelengths (middle panel) the spectrum reveals    
a prominent broad and deep Ly$\beta$ absorption feature, which
is the main feature of the spectrum. In the longer wavelengths
(lower panel) the silicon absorption features Si\,{\sc iii} \& Si\,{\sc iv} 
(around $\lambda \sim 1110-1130$\AA ) are very deep and 
could possibly indicate a high silicon abundance. However, since the
other silicon absorption features are not as pronounced, it is likely that
the deep silicon features are due to incorrect background subtraction.  
Alternatively, there could be some emitting components contributing
some flux around $\lambda \sim 1150-1160$\AA\ . 
The C\,{\sc iii} ($\lambda 1175$) absorption multiplet is also rather deep 
and broad, possibly due to a large velocity broadening. 
The spectrum is rather noisy and of poor quality, making it difficult to model.  

Since the period of the system is not known, we have no way to estimate
its distance. The reddening of the source is also not known but
the galactic reddening in its direction is $E(B-V)=0.09$. 
A reddening of 0.05 will not affect the results by much and therefore, 
in order to assess the effect of the reddening on the results, 
we carry out model fits 
assuming both $E(B-V)=0.0$ and $E(B-V)=0.09$.
Since the system was observed 4 days after an outburst, the flux
is either due to a disk (with a lower accretion rate than the
peak outburst), a heated WD or both. 

\paragraph{Single Disk Models.} 
For a $0.35M_{\odot}$ WD, the best disk model has 
$\dot{M}=10^{-8.5}M_{\odot}$yr$^{-1}$, $i=18^{\circ}$, 
d=4450pc and $\chi^2=0.2127$. As we increase the WD mass, 
the mass accretion rate needed to fit the spectrum decreases
and for   
a $1.2M_{\odot}$ WD, the best disk model has 
$\dot{M}=10^{-10}M_{\odot}$yr$^{-1}$, $i=18^{\circ}$, 
d=2800pc and $\chi^2=0.2138$. The difference in the $\chi^2$ value  
between these best fit models is not significant and we
have therefore no way to find out the mass of the WD 
from these fits. We note that since the system was observed
{\it after} an outburst, it is clear that its mass accretion
rate is much lower than at outburst, 
consistent with the higher WD mass and smaller distance.
Next, we deredden the spectrum assuming $E(B-V)=0.09$
and carry out the same procedure (since the distance we obtained 
is so large, it is likely that the spectrum is affected by
reddening). We find a slightly 
larger mass accretion rate (with about the same distance)
and a $\chi^2$ value larger by about 5-10\%. The smaller
mass accretion rate is obtained for the    
$1.2M_{\odot}$ WD, with  
$\dot{M}=10^{-9.5}M_{\odot}$yr$^{-1}$, $i=18^{\circ}$, 
d=2950pc and $\chi^2=0.2273$.  
None of the disk models provided an  adequate fit to the spectrum
as the deep and broad absorption feature around 1130\AA\ and the 
'hump' feature around 1160\AA\ could not be fit. 

\paragraph{Single WD Models.} 
The lowest mass model ($M=0.4M_{\odot}$) gives a temperature
of 34,000K, a distance of $\sim$3000pc and $\chi^2=0.2115$;
while the largest mass model ($M=1.2M_{\odot}$)  gives a temperature
of 40,000K, a distance of $\sim$1000pc and $\chi^2=0.2125$. 
These models do not provide a better fit than the single disk
models. We then ran an intermediate mass model ($M=0.8M_{\odot}$)
with an increased silicon abundance and found that the fit
could be slightly improved when $Si > 10 \times$ solar. That
best fit model has T=37,000K, d=1700pc and $\chi^2 = 0.2050$
(see Figure 6). 
In order to fit the C\,{\sc iii} ($\lambda 1175$) line profile
all these models have a rotational velocity of about 500km/s. 
Here too we check how a reddening of $E(B-V)=0.09$ 
affects the results and we find no improvement, rather the opposite, 
the $\chi^2$ is larger by about 5-10\%. 
The smallest distance is obtained for the $M=1.2M_{\odot}$ model,
with $T \sim 42,000$K, d=624pc and $\chi^2=0.2290$. 

Lastly, we tried composite WD+disk models but they did not
improve the fit, and because of the unknown mass, distance,
and inclination the results for the composite models were inconclusive.

\subsubsection{NSV 10934/Oct} 
 
NSV10934 is a SU UMa sub-type dwarf nova \citep{kat02,kat04,dow01}.
It shows superoutbursts \citep{kat04} during which
superhumps are detected with a period of 0.07478d.  
Shorter (normal) outbursts occur at intervals of about  40-60 d.  
However, due to the unusual rapid fading of its normal outbursts, 
\citet{kat04} note that it may be an intermediate polar (IP)
candidate showing SU UMa properties. 

NSV 10934 was observed with {\it FUSE} 
on 28 June 2006 (JD2453914.6), 
and on 30 June 2006 (JD2453916.7), 
28 days (or possibly less) after the end of an outburst lasting a week.
At the time of the  {\it FUSE} observation,
the system was in quiescence.  The {\it FUSE} spectrum of NSV 10934
consists of 5 orbits from the first data set and 4 orbits form the
second data set combined together (Figure 7).  
The system was possibly at a magnitude $\sim 15.5-16.0$. 
The flux level of the spectrum is pretty low and the {\it FUSE}
spectrum suffers from a very low S/N. In spite of that we see that 
there is a rather flat continuum with very broad emission lines
from C\,{\sc iii} ($\lambda$977 \& $\lambda$1175) and the O\,{\sc vi}
doublet. 
The broad emission lines are an indication that the system is viewed 
at a high inclination angle, possibly $i \ge 60^{\circ}$ (see e.g.
the broad emission lines of the DN EK TrA, with $i=60^{\circ}$, 
in quiescence \citep{god08a}).  
Since eclipses have not been detected, we can assume $i < 80^{\circ}$. 
The Ly$\beta$ broad absorption feature is not detected, possibly
due to the left wing of the very broad oxygen emission feature. 
However, in that
respect, the continuum is {\it a priori} more characteristic of an
accretion disk seen at a high inclination angle. 
Therefore, in the following we model the {\it FUSE} spectrum
of NSV 10934 with a single WD, a single disk and a WD+disk. 
We also assume two reddening values, E(B-V)=0 and
E(B-V)=0.1 to assess how reddening affects the results. 
Using the period-luminosity relation we find that the distance
to the system is possibly $d \sim 150$pc (see Table 3).

\paragraph{WD models.} 
Since NSV 10934 was observed in quiescence  and has been classified
as a DN, we decided to model its {\it FUSE} spectrum first with
a single WD. For any WD mass between 0.4 to 1.2$M_{\odot}$
there is no synthetic spectrum that can fit the observed spectrum
and the distance at the same time. In order to fit the flux between
1000\AA\ and 1080\AA\  the temperature of the WD has to be of the
order of 25,000K and higher (depending on the mass) and this leads
to a distance well above 200pc. WD models that lead to a distance
of about 200pc have a lower temperature and do not fit the continuum
of the spectrum at all. We then decided to deredden the observed
spectrum assuming E(B-V)=0.1 to see if this improves the modeling and
the best model obtained in this case is for a $1.2M_{\odot}$ WD
with a temperature of 30,000K. This WD model gives a distance
of 216pc and $\chi^2=0.232$ and it is barely better than the WD
models obtained without dereddening the spectrum. We remark here
that the worst models obtained are for the lower WD masses.    

\paragraph{Disk Models.} 
Since the WD models are less than satisfactory, we decided to run
single disk models, first assuming E(B-V)=0.0. 
For $M_{wd} = 0.4 M_{\odot}$ 
the mass accretion rate needed to fit the spectrum
is too large for quiescence and gives a distance far too large.  
For $M_{wd} = 0.8 M_{\odot}$ the best fit model has 
$\dot{M} = 10^{-10}M_{\odot}$yr$^{-1}$ and $i=75^{\circ}$,  and
gives a distance of 185pc and $\chi^2=0.196$. 
The best fit model, however, is obtained  for the largest mass
in the run with  with $M_{wd} = 1.2 M_{\odot}$,  
$\dot{M} = 10^{-10.5}M_{\odot}$yr$^{-1}$ and $i=81^{\circ}$,  and
gives a distance of 144pc and $\chi^2=0.160$ (all shown in Table 4). 
Next we dereddened the spectrum assuming E(B-V)=0.1,   
the best fit model is for    
$M_{wd} = 1.2 M_{\odot}$ with   
$\dot{M} = 10^{-10}M_{\odot}$yr$^{-1}$ and $i=81^{\circ}$,  which 
gives a distance of 187pc and $\chi^2=0.146$. 
This best disk model is shown in Figure 8.
All the models are listed in Table 4.  

At this stage, we tried to add a WD spectral component to the synthetic
disk spectrum but it did not improve the fit. 
We note that with this inclination the WD might not be visible
even with a low mass accretion rate, as the outer part of the
disk might be masking it. The high inclination is also in agreement
with the large broadening of the carbon and oxygen emission features.   

\subsection{Northern Hemisphere Systems} 

\subsubsection{AM Cassiopeiae/258.1928} 
Because of its short period activity and relatively high brightness 
(mag 12.3-15.2 \citep{dow01}) 
AM Cas was discovered long ago by \citet{hof29}. 
Its relation to dwarf novae was suggested 
by \citet{ric61}, who observed a 
9 days outburst cycle and the occurrence of standstills. 
It was classified as a 
Z Camelopardalis system by \citet{not84}.
Optical spectra were later 
obtained by \citet{ric88} and confirmed the classification
of AM Cas as a Z Cam DN as well as its short (mean) outburst cycle period 
(8.2 days).  
The orbital period of the binary system is 0.165 d (=3.96 hr) as
reported by \citet{tay96}, who also obtained 3 optical spectra
$\sim 4500-6300$\AA. One of the spectra is in quiescence with a
flux 
$<  10^{-15}$ergs$~$s$^{-1}$cm$^{-2}$\AA$^{-1}$, and the two other
spectra are in an intermediate state with a maximum flux reaching   
$\sim  10^{-14}$ergs$~$s$^{-1}$cm$^{-2}$\AA$^{-1}$ at 4500\AA\ 
and decreasing toward longer wavelengths.

AM Cas was observed with {\it FUSE}
on 19 October 2006 (JD 2454027.1) and the AAVSO data indicated it
was in standstill at that time with a visual magnitude of 13.8  
(the system can reach 12.3 mag). 
The {\it FUSE} spectrum has a continuum flux level reaching 
$1.5 \times 10^{-13}$ergs$~$s$^{-1}$cm$^{-2}$\AA$^{-1}$ 
(Figure 9),  
which makes AM Cas the brightest FUV target we observed.

The {\it FUSE} data for AM Cas consist of 6 exposures (i.e. orbits). 
The quality
of the spectrum is relatively good and reveals some moderate 
ISM molecular hydrogen 
absorption lines. The C\,{\sc iii} (1175) line might be blue-shifted by 3\AA ,
however this is the region affected by the worm and the  
feature might be an artifact of the worm rather than a true absorption
feature. In the short wavelengths ($< 950$\AA ) the flux does not go to zero
and the continuum there is about 1/3 of the flux at longer 
wavelengths ($>1000$\AA ), indicative of a high temperature. 
On the other side, the Ly$\beta$ is rather deep and wide indicative
of a lower temperature. It is possible that the flux in 
the shorter wavelengths is due to broad emission lines 
from N\,{\sc iv} ($\sim \lambda 923$), S\,{\sc vi} ($\lambda \lambda  
935, 945$), similar to the broad feature around 976\AA - 980\AA\ 
due to C\,{\sc iii} ($\lambda 977$) and N\,{\sc iii} ($\lambda 980$).  
However, there is no sign of emission from the O\,{\sc vi} doublet.  
Another possibility is that of two emitting components (disk+WD), 
one contributing
relatively more flux in the shorter wavelengths, and the other having
a deep Ly$\beta$ profile.  
Since we known AM Cas was observed in an intermediate (standstill) 
state we may expect 
both the WD and the disk to contribute to the flux. 
For the modeling
we assess a distance of 350pc using the period-luminosity relation
(see Table 3). The reddening of AM Cas is not known, however, the  
Galactic reddening in its direction is pretty large ($\sim 0.9$) and it
is therefore possible that AM Cas itself has a significant reddening.  
For this reason, we model the FUSE spectrum of AM Cas assuming 
no reddening and assuming a reddening of 0.2, as this allows us to 
assess how the reddening affects the results.  
We first ran models assuming E(B-V)=0.0.  

\paragraph{Single Disk Models.} 
The single disk models are unable to fit the 
deep Ly$\beta$ feature and the flux in the shorter wavelengths 
at the same time. 
The best models (least $\chi^2$, and consistent with the distance
and outburst state) are as follows. 
For a WD mass $M_{wd}=0.4 M_{\odot}$, 
$\dot{M}= 10^{-8.5} M_{\odot}$yr$^{-1}$, $i \approx 60^{\circ}$,
d=324pc and $\chi^2 = 0.611$; 
as the mass increases to $M_{wd}=0.8 M_{\odot}$, we get  
$\dot{M}= 10^{-9.5} M_{\odot}$yr$^{-1}$, $ i \approx 20^{\circ}$,
d=376pc and $\chi^2 = 0.614$. From the single disk models alone
we infer that the mass of the WD must be $M_{wd} \le 0.8 M_{\odot}$, 
the system must have an inclination $i \le 60^{\circ}$ and 
the mass accretion rate in outburst is     
$\dot{M}= 10^{-9.5}- 10^{-8.5} M_{\odot}$yr$^{-1}$. 

\paragraph{Single WD Models.}
The best fit model was obtained for a $0.4M_{\odot}$
WD with a temperature of 30,000K, a projected rotational velocity
of 1000km/sec, a distance of 307pc, and $\chi^2=0.623$. Models with a larger
mass (and consequently a smaller radius) gave an even smaller distance. 
The single WD model indicates, as the single disk model, a rather small mass
and an intermediate temperature.     

\paragraph{Composite WD+Disk Models.} 
For the composite WD plus disk models we initially chose an
average WD mass of $0.8M_{\odot}$. We find that the least
$\chi^2$ model is for a mass accretion of 
$2 \times 10^{-10}M_{\odot}$/yr, an inclination $i=18^{\circ}$, 
a WD temperature $T=36,000$K with a projected rotational
velocity $V_{rot} \sin{i}$=500km/s. This gives a distance of
373pc,  $\chi^2=0.563$, with the WD contributing 41\% of the
flux and the disk 59\%. This model, though better than
the single WD and single disk models, also does not fit 
too well the high flux in the lower wavelengths. 
This model is shown in Figure 10.
In order to fit the lower wavelength region of the
spectrum we have to increase the temperature of the WD to about 40,000K.
This in turn increases the distance to $d>400$pc and $\chi^2>0.57$.  
We note that performing the same WD+disk composite model fit with 
a lower mass gave about the same results but with a larger distance, 
while a larger mass gave a smaller distance. However, we do not 
consider these models here as we used a distance in agreement
with the distance obtained using the period-mangitude relations 
of \citet{war87} and \citet{har04}, namely $d\approx 350$pc (see
Table 3).   

\paragraph{E(B-V)=0.2}   
We performed the same model fits.  
The best fit model is a WD+disk
composite model with $M=0.55M_{\odot}$,  
$\dot{M}= 3 \times 10^{-9}M_{\odot}$/yr, an inclination $i=18^{\circ}$, 
a WD temperature $T=35,000$K with a projected rotational
velocity $V_{rot} \sin{i}$=400km/s. The distance obtained is
331pc,  $\chi^2=0.538$, with the WD contributing only 11\% of the
flux and the disk 89\% (Figure 11).  
This model does better at fitting the spectrum  
both at the shorter and longer wavelengths. However,
it fits the Ly$\beta$ region a little less well.  The higher mass
models do not perform as well and give a distance $d<250$pc.

\subsubsection{FO Persei/22.1939}

FO Per has a mean outburst cycle of about 10 days 
\citep{how76,ges78} and belongs to those DN with a particularly
high outburst rate. 
Its magnitude varies between 11.8 and 16.2 \citep{dow01}.  
It was observed in the optical by \citet{bru87,bru89} but these and 
more recent observations have not been able to unambiguously determine
the orbital period of the system: it is not clear whether the
period is 3.52h or 4.13h \citep{she07}. 
However, with these values we assess the distance to FO Per  
using the absolute outburst magnitude/Period relation 
\citep{war87,har04} (see Table 3) and we obtain $d\approx 270$pc. 

FO Per was observed with {\it FUSE} on 11 Feb 2007 (JD2454142.1) and
the AAVSO data indicate it was caught in an intermediate  state
of brightness at 13.4 mag.
The {\it FUSE} spectrum of FO Per is of moderately good quality
exhibiting both interstellar (molecular hydrogen) and intrinsic
absorption features (see Figure 12).
There is definitely
more flux in the longer wavelengths: this is a sign of either a strong
reddening or a moderate temperature.  Except for air glow, there are no
emission lines. 
 
Even though FO Per was caught with  
{\it FUSE} while reaching 13.4 mag (brighter than AM Cas at 13.8 mag),
it has a continuum FUV flux level 3 times lower than that of AM Cas: 
namely, at the time of the observations FO Per was {\it redder} than AM Cas. 
This is consistent with the fact that the galactic reddening in the
direction of FO Per is extremely large, namely E(B-V)= 1.78, 
as the dust reddening
in the direction of the constellation of Perseus is quite large. 
To estimate the reddening of FO Per, we note (Table 3) that 
TZ Per has a known reddening of 0.27 at a distance of about 500pc,
with a galactic reddening of 0.8. Since FO Per is about half
this distance but with a galactic reddening about twice as large,
we assume that the reddening of FO Per is about 0.3.  
We carry out simulations assuming 
both E(B-V)=0.0 and E(B-V)=0.3, to assess the effect
of the reddening on the results. 

\paragraph{E(B-V)=0.0} 
For the single disk models 
we find little agreement between the models and
the observed spectra. 
Assuming $M=0.4M_{\odot}$ we find a mass accretion rate 
$\dot{M}= 10^{-8.5}M_{\odot}$/yr; while for the larger
mass models $M=1.2M_{\odot}$ the mass accretion rate decreases to
a quiescent value $10^{-10.5}M_{\odot}$/yr. 
Next we try the single WD models and find that they do not provide
a better fit.
The lower mass models
have a temperature of $\sim 25,000$K, and the higher mass models
have a temperature of up to $45,000$K. Because of the rather poor
fit the rotational velocity does not influence the results, and we
find that the 200km/s models are about as good as the 400km/s models.  
We then run the combined WD+disk model-fits and find
the best fit to be for a $0.4M_{\odot}$ WD mass, with 
$T_{wd}=21,000$K, $V_{rot} \sin{i}=$200km/s, 
$i=75^{\circ}$, $\dot{M} = 10^{-8.5}M_{\odot}$yr$^{-1}$,
d=291pc, and the least $\chi^2$ is $0.258$.
The WD contributes 29\% of the FUV flux while the disk
contributes 71\%.  This model is shown in Figure 13.

\paragraph{E(B-V)=0.3} 
The best-fit single disk model has 
$M_{wd}=1M_{\odot}$, $i=18^{\circ}$, $\dot{M} = 10^{-9}M_{\odot}$yr$^{-1}$,
and gives a distance of 281pc and a much smaller $\chi^2$, namely 0.172.
The single WD model gives a smaller $\chi^2$ but also a much
smaller distance. For a $0.4M_{\odot}$ WD mass, the best fit has a
temperature $T_{wd}=35,000$K, a rotational velocity of 200km/s, 
and gives a distance of 135pc and $\chi^2=0.152$. For larger WD
masses we find an even smaller distance, and we therefore disregard
the single WD models. 
We then try the combined WD+disk models and find
that a valid distance  of 254pc is obtained for 
a $0.4M_{\odot}$ WD mass, with 
$T_{wd}=40,000$K, $V_{rot} \sin{i}=$200km/s, 
$i=18^{\circ}$, $\dot{M} = 10^{-8.5}M_{\odot}$yr$^{-1}$,
and this best model gives the least $\chi^2$ of all $0.146$. 
The WD contributes 52\% of the FUV flux while the disk
contributes the remaining 48\%.  
This model is shown in Figure 14.

\subsubsection{ES Draconis/Dra3/PG 1524+622} 

ES Dra has a magnitude varying between 13.9 to 16.3 \citep{and91,dow01}.
First suspected to be a CV by \citet{gre86}, 
it is classified as a SU UMa system in \citep{dow01} and the non-detection of
prominent superhumps indicates that it might have a low inclination 
\citep{and03}. 
However, its orbital period of 4.29 hr \citep{ring94} makes
its classification as an SU UMa unlikely.     
ES Dra appears to sit about 1 mag below outburst for months,
this is more characteristic of Z Cam systems. We therefore tentatively
classify ES Dra as a Z Cam, this classification is also consistent
with its location above the period gap. 

ES Dra was observed with {\it FUSE} on 19 Nov 2006 (JD2454059). 
However, no AAVSO data were collected around the period of the 
actual {\it FUSE} observation. From the {\it FUSE} continuum flux level 
($2 \times 10^{-14}$ergs$~$s$^{-1}$cm$^{-2}$\AA$^{-1}$) it seems that
ES Dra must have been in an intermediate (standstill) state         
possibly at 15.0-15.4 mag, where the heated WD might be the main 
component of the FUV continuum. The spectrum of ES Dra is shown
in Figure 15.

The C\,{\sc iii} (1175) multiplet possibly exhibits a P-Cygni profile,
where the broad absorption line is blue-shifted by about 3\AA ,
and the broad emission line  is red-shifted by about the same amount. 
This corresponds to a velocity of about $\sim 750$km/s.  
Such a high velocity is probably originating close to the
WD/inner disk.
The C\,{\sc iii} (977) broad emission line is also red-shifted
by at least 2\AA  . 
The two broad S\,{\sc iv} absorption lines ($\sim$1063+1073) 
are blue shifted by 2\AA , while   
the Si\,{\sc iii}+{\sc iv} ($\sim$1110+1125) 
broad absorption features also show a slight blue-shift.   
It is clear that the C\,{\sc iii} and S\,{\sc iv} lines are forming in 
an expanding shell or a corona, 
while the Si absorption feature also forms in the disk 
and atmosphere of the WD. 
We therefore decided to  mask the C\,{\sc iii} and S\,{\sc iv} lines 
in the model fitting, while keeping the Si lines. 

We assume that the reddening toward ES Dra is negligible as the
Galactic (maximum) reddening in that direction is itself very small: 0.03. 
The distance to the system inferred from the period-magnitude
relation is 770pc (see Table 3), and this is the distance
we use in the modeling.   

\paragraph{Single Disk Models.} 
The lowest $\chi^2$ model in best agreement
with a distance of 770pc is for $M=1.2M_{\odot}$, $\dot{M} = 10^{-10} 
M_{\odot}$yr$^{-1}$, which gives a distance of 1035pc and $\chi^2 > 0.57$.
The fit can be improved by changing the abundances, namely  
the silicon and sulfur broad absorption features are
best fit when $Si \approx 10 \times$ solar and $S \approx 50 \times$ solar.
The fit is further improved (lower $\chi^2$) by blue-shifting
the entire synthetic spectrum by 1.0\AA .  This best single
disk model gave a distance 1754pc, and $\chi^2=0.4280$.
This model is shown in Figure 16. 

\paragraph{Single White Dwarf Models.} 
We tried to fit both the
distance (770pc) and the higher order of the Ly series. We first tried solar
abundances models and then we increased the silicon and sulfur
abundances to better fit the silicon and sulfur absorption features. 
The best model (Figure 17) has $T=35,000$K, 
$log(g)=7.8$ (corresponding to $M_{wd} \approx 0.58M_{\odot}$), 
a projected rotational velocity $v_{rot} \sin{i} = 700$km/s, .  
Here too, we find that the best fit is obtained with 
$Si \approx 10 \times$ solar, $S \approx 50 \times$ solar and a 
blue-shift of 1.3\AA . 
The distance obtained is 773pc and $\chi^2 = 0.4277$. 
From the point of view of the least $\chi^2$, this model is as good
as the least $\chi^2$ single disk model, however the distance is in
much better agreement with the estimate of 770pc (Table 3). 

We find that WD+Disk composite models do not improve
the fit, but rather the opposite. 
 
\section{Summary}
 
We report the results of a spectroscopic analysis of
a {\it FUSE} survey of high-declination DNs, including five 
Southern Hemisphere objects and three Northern Hemisphere objects. 
In order to model these spectra we assessed the distance and
reddening of these objects and carried out a synthetic spectral
fit to the observed spectra. 

The low S/N in some of the spectra prevented us from obtaining
robust results: 
for AQ Men and V433 Ara, we were only able to assess a
lower limit
for the WD temperature or mass accretion rate.  
Though we were unable to assess their distance, 
for HP Nor we found a WD temperature of about 40,000K and
rotational velocity of 1000km/s, 
and for DT Aps we found a WD temperature of at least
34,000K and a rotational velocity of 500km/s.   

For the 4 objects with relatively good spectra, we 
were able to obtain more reliable results. 
For NSV 10934 we find a low mass accretion rate,
high inclination angle, and a large WD mass. 
For AM Cas the best fit model is a WD+disk
composite model with a low inclination,
a WD temperature T=35,000K, a mass $0.55 M_{\odot}$, 
and a mass accretion rate 
of $3 \times 10^{-9} M_{\odot}$/yr. 
For FO Per the best fit is also a WD+ disk
composite model giving $M=0.4M_{\odot}$, 
T=40,000K, a low inclination and a mass accretion
rate consistent with outburst ($10^{-8.5} M_{\odot}$/yr). 
And for ES Dra the spectrum is best fit with a
single WD stellar atmosphere with T=35,000K,
$M=0.58 M_{\odot}$, and a projected rotational velocity
of 700km/s with over-abundant silicon and sulfur.  

For 3 objects above the period gap (AM Cas, FO
Per and ES Dra) we have found
a WD temperature during an active state (intermediate-to-outburst)  
reaching 35,000-40,000K. These must be regarded as 
an upper limit as the WD was heated due to 
on-going (or recent) accretion. These temperatures are similar
to the temperature of RX And on the rise to outburst 
(39,000K) and in decline from outburst (45,000K), while in 
quiescence its temperature is 34,500K \citep{sep02}.  
It is therefore likely that the true temperatures of AM Cas, FO Per
and ES Dra in quiescence are around 30,000-35,000K, and these 
objects are therefore near the average CV WD's $T_{eff}$
above the period gap ($\sim$30,000K), similar to U Gem, 
SS Aur and RX And.  

We have recapitulated the effective quiescent CV WD temperatures as 
found in the literature in Table 5. The references are listed in the last
column. 
Similar tables are given
in the literature \citep{tow09} and sometimes reflect slightly 
different values, this is due to the facts that 
(i) for some of the systems more than one model (and therefore more than
one WD temperature) was given and we list here the best fit in which the
WD contributes more than 50 percent of the flux; 
(ii) we also list the most recent synthetic spectral fits obtained by our
own group. 
For WZ Sge and VW Hyi, 
a theoretical cooling curve of the WD was computed to fit the observed 
temperature decline in these systems:   
the quiescent temperature was the asymptotic value approached as $t\rightarrow
\infty$.  
The temperatures listed in Table 5 are depicted in Figure 18, 
from which there seems to be a clear separation between
Polars and DNs, both above and below the gap 
\citep{ara03}. 
VY Scl systems also seem
definitely hotter than DNs. It is evident from this figure that more
data points are badly needed for Z Cam systems as well as for all 
the CV sub-types
above the gap, except maybe for U Gem's. The data points from the present work
have been added as downward pointing arrows, as they are only upper limits.    
Since the mass and distance is not known with accuracy for many of the systems
listed in Table 5, it is very likely that some of the temperatures in Table 5
will change once parallaxes and/or masses are obtained for these systems.   
The standard model above the gap (traditional magnetic braking \citet{how01}) 
is between the dotted lines. 

The recent work of \citet{tow09} recapitulates and analyzes the 
effective temperatures of the WD in CVs, 
therefore, we wish here only
to emphasize the (small) differences that we obtain above the
period gap when considering 
the values given in Table 5. We include in the present work more data
points for the VY Scl systems (BB Dor, V794 Aql, VY Scl) and DNs, 
and we also draw systems with $P>5h$.   
Since we use the same data points for the magnetic systems, we
agree that gravitational radiation can account for the WD $T_{eff}$
(and therefore $\dot{M}$) of polars both above and below the gap. 
Above the gap we also agree that the majority of DNs have a 
temperature (mass accretion rate)
lower than expected by the standard theory, and that
NL VY systems have a temperature higher than expected from the standard
theory.  
However, we find that the traditional magnetic  braking
could explain the temperature (mass accretion rate) of some systems 
(TU Men, BB Dor, SS Aur, U Gem) above the period gap. 

Overall it seems that the standard model (and any modified
model: see \citet{tow09} for a discussion) does not agree with the
non-magnetic CVs above the period gap. While it is difficult to explain
the discrepancy above the gap 
between the theory and observations for the DN systems, 
a higher than expected temperature for the NL VY can be accounted for
with a higher mass accretion rate $<\dot{M}>$. 

\acknowledgments

PG wishes to thank Mario Livio, for his kind hospitality at the
Space Telescope Science Institute
where part of this work was carried out.
We wish to thank the members of the American Association of
Variable Star Observers (AAVSO), the Austral Variable Star Observer
Network (AVSON), and the Center for Backyard Astrophysics (CBA)
for providing us with optical monitoring of Northern and
Southern hemisphere systems, as well as optical archival data
on these systems.
This research was based on observations made with the
NASA-CNES-CSA Far Ultraviolet Spectroscopic Explorer.
{\it{FUSE}} is operated for NASA by the Johns Hopkins University under
NASA contract NAS5-32985. This work was supported by the National
Aeronautic and Space Administration (NASA)
under FUSE Cycle 7 (Guest Investigator Program) grant NNX06AD28G
and supported in part by NSF grant AST0807892,
both to Villanova University.
This publication also makes use of the data products from the Two Micron
All Sky Survey, which is a joint project of the University of
Massachusetts and the Infrared Processing and Analysis Center/California
Institute of Technology, funded by the National Aeronautics and
Space Administration and the National Science Foundation.

\clearpage 

\setlength{\hoffset}{-15mm} 

\begin{deluxetable}{lllcclc}
\tablewidth{0pc}
\tablecaption{System Parameters of the {\it FUSE} Targets}
\tablehead{
Name$^a$ &Type     & P$_{orb}$ & $i$   &V$_{max}$&V$_{min}$ & Priority  \\
         &         & (hr)    & (deg) &         &          &          
} 
\startdata 
VW Tuc   &UG       & ---       & $>20$? &   15.2  &$<$ 16.5 & 2      \\
AQ Men   &UG/NL?   & 3.40      & eclipsing& 14    & 15      & 1      \\
HP Nor   &   ZC    & ---       & ---    &   12.8  & 16.4    & 2      \\
IK Nor   &UG       & ---       & ---    &   12.9  & 16.3    & 2      \\
V663 Ara &   SU    & ---       & ---    &   15.9  &$<$ 16.3 & 2      \\
V433 Ara &   ZC/NL?& 4.70      & ---    &   14.8  &$<$ 16.3 & 2      \\
V499 Ara &UG       & ---       & ---    &   14.8  &$<$ 16.2 & 2      \\
DT Aps   &   SS    & ---       & ---    &   14.4  &$<$ 15.8 & 1      \\
NSV 10934&   SU    & 1.74      & ---    &   11.2  & 15.0    & 1      \\
AM Cas   &   ZC    & 3.96      & ---    &   12.3  & 15.2    & 1      \\
FO Per   &UG       & 3.52/4.13   &--- &   11.8  & 16.2    & 1      \\
ES Dra   &  ZC     & 4.29      &  low?  &   13.9  & 16.3    & 1      \\ 
\enddata 
\tablenotetext{a}{The Targets are listed in order of ascending RA, Southern 
Hemisphere objects followed by Norther Hemisphere objects.} 
\end{deluxetable} 

\clearpage

\begin{deluxetable}{lcrllcccl}
\tablewidth{0pc}
\tablecaption{FUSE Observations Log}
\tablehead{
System  & Date   & Exp.time & Dataset &Aperture&Operation& V       & State        & Data    \\ 
Name    &(dd/mm/yy) & (sec)$~~~$&     &       & Mode   &           &                         & Quality  
}
\startdata 
VW Tuc  & 11-06-06  & 8,835 & G9250101 & LWRS  & TTAG  & 16-18?    & quiescence   & unusable \\
AQ Men  & 22-11-06  & 22,661& G9250201 & LWRS  & TTAG  & 14-15     & usual state  & poor     \\ 
HP Nor  & 13-04-07  & 3,952 & G9250601 & LWRS  & TTAG  & 15.0      & low          & poor     \\ 
IK Nor  & 13-04-07  & 7,947 & G9250701 & LWRS  & TTAG  & ---       &  ---         & unusable \\ 
V663 Ara & 12-04-07 & 4,807 & G9250802 & LWRS  & TTAG  & ---       &  ---         & unusable \\ 
         & 24-04-06 & 9,785 & G9250801 & LWRS  & TTAG  & ---       &  ---         & unusable \\ 
V433 Ara & 18-02-07 & 7,009 & G9250901 & LWRS  & TTAG  & ---       &  ---         & unusable \\ 
V499 Ara & 22-06-06 & 17,988& G9251001 & LWRS  & TTAG  & ---       &  ---         & unusable \\ 
DT Aps   & 03-05-06 & 8,053 & G9251102 & LWRS  & TTAG  & ---       & decline/low  & unusable \\ 
         & 28-04-06 & 68,245& G9251101 & LWRS  & TTAG  & ---       & active       & poor  \\ 
NSV 10934 &28-06-06 & 11,956& G9251201 & LWRS  & TTAG  & ---       &  ---         & poor  \\ 
          &30-06-06 & 14,177& G9251202 & LWRS  & TTAG  & ---       &  ---         & poor  \\
AM Cas  & 19-10-06  & 12,909& G9251402 & LWRS  & TTAG  & 13.3-13.8 & intermediate & good  \\ 
FO Per  & 11-02-07  & 2,716 & G9251501 & LWRS  & TTAG  & 13.5      & intermediate & good  \\ 
ES Dra  & 19-11-06  & 24,554& G9251601 & LWRS  & TTAG  & 15.1-15.4 & intermediate & good  \\ 
\enddata 
\end{deluxetable} 

\clearpage 

\begin{deluxetable}{llclccccc}
\tablewidth{0pc}
\tablecaption{Parameters Adopted for the Modeling} 
\tablehead{
System&Period& Gal.Lat. & $E_{B-V}^{Gal}$&$E_{B-V}^{adp}$& $d_w$ & $d_h$ & $d_k$    &$d$   \\
Name  &(hr)  &  (deg)   &    &               & (pc)  &  (pc) & (pc)     & (pc)    
} 
\startdata 
HP Nor   & ---      & -3   & 0.63   &   0.20        & ---   & ---   & ---      &  ---   \\
DT Aps   & ---      & -20  & 0.09   &   0.09        & ---   & ---   & ---      &  ---   \\
NSV 10934& 1.74     & -27  & 0.16   &   0.10        & 151   & 155   & 117-219  &   150  \\
AQ Men   & 3.40     & -31  & 0.18   &   0.10        & 664   & 752   & 201-377  &   710  \\
FO Per   & 3.52     &  0   & 1.78   &   0.30        & 244   & 279   & 182-342  &   260  \\
         & 4.13     &  0   & 1.78   &   0.30        & 262   & 311   & 227-311  &   285  \\
AM Cas   & 3.96     &  9   & 0.89   &   0.20        & 324   & 380   & 153-287  &   350  \\
ES Dra   & 4.29     & 47   & 0.02   &   0.00        & 702   & 840   & 540-1016 &   770  \\
V433 Ara & 4.70     & -9   & 0.19   &   0.10        & 1114  & 1368  & 747-1411 &  1200  \\
TZ Per    & 6.31    & -3   & 0.78   &   0.27        & 424   & 575   &  ---     & 500   \\  
\enddata 
\tablenotetext{-}{
The systems are listed in order of increasing binary period.
In column (3) we list the Galactic Latitude.  
In column (4) we list the Galactic reddening which is anti-correlated
to the Galactic Latitude.  
as inferred from the 100$\micron$ dust emission map \citep{sch98}.
In column (5) we list the reddening we used in our modeling.  
In column (6) we list the distance assessed using the period-magnitude 
correlation given by \citet{war87}.  
In column (7) we list the distance assessed using the period-magnitude 
correlation given by \citet{har04}. 
In column (8) we list the distance estimated using the secondary 
IR emission technique of \citet{kni06,kni07}. 
In the last column we list the distance we used in our modeling.  
TZ Per was not observed but has been added for comparison.
}
\end{deluxetable} 

\clearpage 

\begin{deluxetable}{lcccccccccc}
\tablewidth{0pc}
\tablecaption{Synthetic Spectra}
\tablehead{
System & $M_{wd}$&$T_{wd}$&$V_{rot}sin{i}$& $i$ & $Log(\dot{M})$  &WD/Disk& d &$\chi^2_{\nu}$&$E_{B-V}$ &  Fig. \\
Name &$(M_{\odot})$&($10^3$K)&(km/s) & (deg) & ($M_{\odot}$/yr) &(\%) &(pc)&           &          &    \\
}
\startdata
HP Nor & 0.4    & 34    &  1000     &  ---  & ---     & WD     &1134  & 0.105  & 0.00 &    \\
       & 0.8    & 37    &  1000     &  ---  & ---     & WD     & 648  & 0.107  & 0.00 &    \\
       & 1.2    & 40    &  1000     &  ---  & ---     & WD     & 381  & 0.105  & 0.00 &    \\
       & 0.55   & ---   &  ---      &  60   & -8.5    & Disk   &1448  & 0.105  & 0.00 &    \\
       & 0.8    & ---   &  ---      &$<41$  & -9.0    & Disk   &1603  & 0.104  & 0.00 &    \\
       & 1.2    & ---   &  ---      & 8     & -10.0   & Disk   &1021  & 0.101  & 0.00 &    \\
       & 0.4    & 40    &  1000     &  ---  & ---     & WD     & 530  & 0.096  & 0.20 &        \\
       & 0.8    & 43    &  1000     &  ---  & ---     & WD     & 265  & 0.096  & 0.20 &     3  \\
       & 1.2    & 46    &  1000     &  ---  & ---     & WD     & 148  & 0.096  & 0.20 &        \\
       & 0.55   & ---   &  ---      &  18   & -8.0    & Disk   &1060  & 0.097  & 0.20 &        \\
       & 0.8    & ---   &  ---      &  18   & -8.0    & Disk   &1800  & 0.093  & 0.20 &        \\
       & 1.2    & ---   &  ---      &  18   & -8.5    & Disk   &1515  & 0.095  & 0.20 &        \\
AQ Men &  0.4   &  24   & ---       &  ---  &  ---    & WD     & 710  &  ---   & 0.00 &     \\
       &  1.1   &  36   & ---       &  ---  &  ---    & WD     & 710  &  ---   & 0.00 &     \\
       & 0.35   & ---   & ---       &   81  & -8.75   & Disk   & 710  &  ---   & 0.00 &     \\
       & 1.20   & ---   & ---       &   81  & -9.5    & Disk   & 710  &  ---   & 0.00 &     \\
       &  0.4   &  28.5 & ---       &  ---  &  ---    & WD     & 710  &  ---   & 0.10 &     \\
       &  1.1   &  60   & ---       &  ---  &  ---    & WD     & 710  &  ---   & 0.10 &     \\
       & 0.35   & ---   & ---       &   81  & -7.9    & Disk   & 710  &  ---   & 0.10 &     \\
       & 1.20   & ---   & ---       &   81  & -8.9    & Disk   & 710  &  ---   & 0.10 &     \\
V433 Ara &  0.4   & ---   & ---     &  81   & -8.5    & Disk   & 1200 & ---    & 0.00   &   \\  
         &  0.4   & ---   & ---     &  81   & -8.0    & Disk   & 1200 & ---    & 0.10   &   \\  
         &  1.2   & ---   & ---     &  81   & -9.5    & Disk   & 1200 & ---    & 0.00   &   \\  
         &  1.2   & ---   & ---     &  81   & -9.0    & Disk   & 1200 & ---    & 0.10   &   \\  
         &  0.4   & 24    & ---     &  ---  & ---     & WD     & 1200 & ---    & 0.00  &     \\
         &  0.4   & 28    & ---     &  ---  & ---     & WD     & 1200 & ---    & 0.10  &     \\
         &  0.8   & 30    & ---     &  ---  & ---     & WD     & 1200 & ---    & 0.00  &     \\
         &  0.8   & 40    & ---     &  ---  & ---     & WD     & 1200 & ---    & 0.10  &     \\
         &  1.2   & 40    & ---     &  ---  & ---     & WD     & 1200 & ---    & 0.00  &     \\
         &  1.2   & 75    & ---     &  ---  & ---     & WD     & 1200 & ---    & 0.10  &     \\
DT Aps &$\le 1.2 $& --- & ---       &  18  &$\ge-10$ & Disk  &$\ge 2800$&0.213  & 0.00 &    \\
       &$\le 1.2 $& --- & ---       &  18  &$\ge-9.5$ & Disk  &$\ge 2950$&0.223  & 0.09 &    \\
       &$\ge 0.4$&$\ge34$& 500      &  --- &  ---     & WD    &$\ge 1000$&0.212  & 0.00 &    \\
       &  0.8    &  37   & 500      &  --- &  ---     & WD    &  1700    &0.205  & 0.00 &   8    \\ 
       &$\ge 0.4$&$\ge36$& 500      &  --- &  ---     & WD    &$\ge 625$&0.225  & 0.09 &    \\
NSV 10934 &0.8  & ---   &  ---      &  75   &  -10    & Disk   & 185  &  0.196 & 0.00 &     \\
          &1.2  & ---   &  ---      &  81   &  -10.5  & Disk   & 144  &  0.160 & 0.00 &     \\
          &0.4  & ---   &  ---      &  75   &  -9     & Disk   & 179  &  0.202 & 0.10 &     \\
          &0.8  & ---   &  ---      &  80   &  -9.5   & Disk   & 157  &  0.172 & 0.10 &     \\
          &1.2  & ---   &  ---      &  81   &  -10    & Disk   & 187  &  0.146 & 0.10 &   11  \\ 
          &1.2  & 30    &  ---      &  ---  &  ---    & WD     & 216  &  0.232 & 0.10 &     \\
AM Cas &$\le0.8$& ---   & ---       &$\le 60$&$\ge-9.5$& Disk  & 350  &  0.611 & 0.00 &     \\
       & 0.40   & 30    & 1000      &  ---  & ---     & WD     & 307  &  0.623 & 0.00 &     \\
       & 0.80   & 36    &  500      &  18   & $-9.7$  & 41/59  & 373  &  0.563 & 0.00 &   14 \\ 
       & 0.55   & 35    &  400      &  18   & $-8.5$  & 11/89  & 331  &  0.538 & 0.20 &   15  \\ 
FO Per & 0.40   & ---   & ---       &   75  & -8.5    & Disk   & 251  & 0.296  & 0.00 &     \\
       & 0.80   & ---   & ---       &   60  & -9.5    & Disk   & 295  & 0.313  & 0.00 &     \\
       &$\ge$0.40 &$\ge$25  & 400   &  ---  & ---     & WD     & 266  & $\ge$0.312 & 0.00 &     \\
       & 0.40   & 21    & 200       &  75   & -8.5    & 29/71  & 291  & 0.258  & 0.00 &  18 \\ 
       & 1.00   & ---   & ---       &   18  & -9.0    & Disk   & 281  & 0.172  & 0.30 &     \\
       &$\ge0.40$ & 35  & 200       &  ---  & ---     & WD     &$\le$135 & 0.152 & 0.30 &     \\
       &$\ge$0.40 & 40  & 200       &  18   &$\le$-8.5 & 52/48  &$\le$254 & 0.146 & 0.30 &  19 \\ 
ES Dra & 1.20   & ---   & ---       &   5   & -9.5    & Disk   & 770  & 0.4280 & 0.00 &    22   \\ 
       & 0.58   & 35    & 700       &  ---  & ---     & WD     & 770  & 0.4277 & 0.00 &    23  \\ 
\enddata
\end{deluxetable}

\clearpage 

\begin{deluxetable}{llrrl}
\tablewidth{0pc}
\tablecaption{CV WD Temperatures in Quiescence} 
\tablehead{
Name$^a$ &Type     & P$_{orb}$ & $T_{wd}$   & Reference   \\
         &         & (min)     & (K)        &                  
} 
\startdata 
GW Lib   & DN SU   & 76.8      & 13,300     & \citet{szk02a}  \\
BW Scl   & DN SU   & 78.2      & 14,800     & \citet{gan04}  \\
LL And   & DN SU   & 79.2      & 14,300     & \citet{how02}  \\ 
EF Eri   & NL AM   & 81.0      &  9,500     & \citet{beu00}  \\ 
WZ Sge   & DN SU   & 81.6      & 13,500     & \citet{god06b}  \\ 
AL Com   & DN SU   & 81.6      & 16,300     & \citet{szk03}  \\ 
SW UMa   & DN SU   & 81.8      & 14,000     & \citet{gan04,urb06}  \\ 
HV Vir   & DN SU   & 83.5      & 13,300     & \citet{szk02b}  \\ 
WX Cet   & DN SU   & 83.9      & 14,500     & \citet{sio03,urb06}  \\ 
T Leo    & DN SU   &  84.7     & 16,000     & \citet{ham04}  \\
EG Cnc   & DN SU   & 86.4      & 13,300     & \citet{szk02b}  \\ 
DP Leo   & NL AM   & 89.8      & 13,500     & \citet{sch02}  \\ 
V347 Pav & NL AM   & 90.0      & 12,300     & \citet{ara05}  \\ 
BC UMa   & DN SU   & 90.2      & 15,200     & \citet{gan04,urb06}  \\ 
VY Aqr   & DN SU   & 90.8      & 14,000     & \citet{sio03,urb06}  \\ 
EK TrA   & DN SU   & 91.6      & 17,000     & \citet{god08a}  \\ 
BZ UMa   & DN SU   & 97.9      & 17,500     & \citet{god07b,urb06}  \\ 
VV Pup   & NL AM   & 100.4     & 12,100     & \citet{ara05}  \\ 
V834 Cen & NL AM   & 101.5     & 14,200     & \citet{ara05}  \\ 
HT Cas   & DN SU   & 106.1     & 18,000     & \citet{urb06}  \\
VW Hyi   & DN SU   & 106.9     & 19,000     & \citet{sio09}  \\ 
CU Vel   & DN SU   & 113.0     & 18,500     & \citet{gan99}  \\ 
BL Hyi   & NL AM   & 113.6     & 13,100     & \citet{ara05}  \\ 
MR Ser   & NL AM   & 113.5     & 14,000     & \citet{ara05}  \\ 
ST LMi   & NL AM   & 113.9     & 10,800     & \citet{ara05}  \\ 
EF Peg   & DN SU   & 123.0     & 16,600     & \citet{how02}  \\ 
HU Aqr   & NL AM   & 125.0     & 14,000     & \citet{gans99}  \\ 
UV Per   & DN SU   & 125.0     & 20,000     & \citet{urb06}  \\
QS Tel   & NL AM   & 139.9     & 17,500     & \citet{ros01}  \\ 
TU Men   & DN UG   & 168.8     & 28,000     & \citet{sio08}  \\
AM Her   & NL AM   & 185,6     & 19,800     & \citet{gan06}  \\ 
MV Lyr   & NL VY   & 191.0     & 47,000     & \citet{hoa04}  \\ 
DW UMa   & NL SW   & 196.7     & 50,000     & \citet{ara03}  \\ 
TT Ari   & NL VY   & 198.0     & 39,000     & \citet{gea99}  \\ 
BB Dor   & NL VY   & 214.8     & 32,000     & \citet{god08b}  \\
V794 Aql & NL VY   & 220.8     & 47,000     & \citet{god07}  \\
UU Aql   & DN UG   & 235.5     & 27,000     & \citet{sio07}  \\
X Leo    & DN UG   & 237.0     & 33,000     & \citet{urb06}  \\
AM Cas   & DN ZC   & 237.6     &$<$35,000   & this work   \\ 
VY Scl   & NL VY   & 239.3     & 45,000     & \citet{ham08} \\ 
FO Per   & DN UG   & 229.5$^{a}$ &$<$40,000  & this work   \\ 
V1043 Cen& NL AM   & 251.4     & 15,000     & \citet{gan00}  \\ 
WW Cet   & DN      & 253.1     & 26,000     & \citet{god06a}  \\
U Gem    & DN UG   & 254.7     & 31,000     & \citet{sio98,lon99,urb06}  \\ 
ES Dra   & DN ZC   & 257.4     &$<$35,000   & this work   \\ 
SS Aur   & DN UG   & 263.2     & 34,000     & \citet{god08a}  \\ 
V895 Cen & NL AM   & 285.9     & 13,800     & \citet{ara05}  \\ 
RX And   & DN ZC   & 302.2     & 34,000     & \citet{sio01,urb06}  \\ 
TT Crt   & DN UG   & 386.4     & 30,000     & \citet{sio08,urb06}  \\
SS Cyg   & DN UG   & 396.2     & 55,000     & \citet{sio07} \\ 
Z Cam    & DN ZC   & 417.4     & 57,000     & \citet{har05}  \\ 
RU Peg   & DN UG   & 539.4     & 70,000     & \citet{god08a}  \\ 
EY Cyg   & DN UG   & 661.4     & 30,000     & \citet{god08a}  \\ 
V422 Cen & DN UG   & 662.4     & 47,000     & \citet{sio08}  \\
BV Cen   & DN UG   & 878.6     & 40,000     & \citet{sio07}  \\
\enddata 
\tablenotetext{a}{   
For FO Per it is not known whether the period is 
211.2min or 247.8min, we therefore took the intermediate value 229.5min
for practical reasons.}  
\end{deluxetable} 

\clearpage 
\begin{figure}
\vspace{-5.cm} 
\plotone{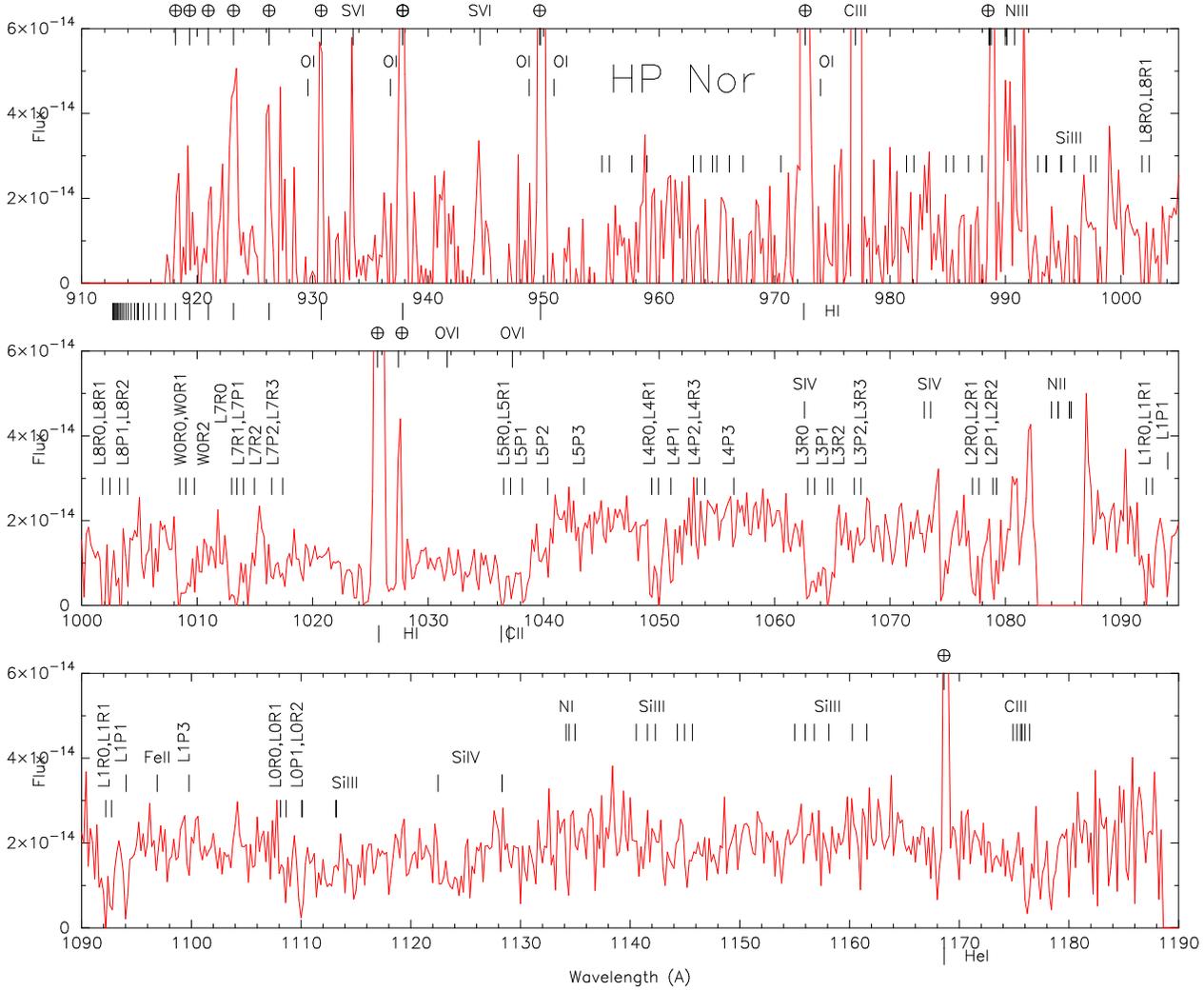} 
\caption{The {\it FUSE} spectrum of HP Nor. The sharp emission lines
(including S\,{\sc vi} 944 \AA) 
are due to geo- and helio-coronal contamination. 
The ISM molecular features are denoted for
clarity. The C\,{\sc iii} (1175\AA) absorption feature is identified
possibly with a red  shift of $\sim 2$\AA.
The S\,{\sc iv} (1073\AA), Si\,{\sc iii} (1113\AA), Si\,{\sc iv} 
(1122 \& 1128\AA) lines could also be accounted for with a  red-shift
of $\sim 2$\AA.  
Other metal lines have been marked but are not detected.}  
\end{figure}

\clearpage 
\begin{figure}
\vspace{-5.cm} 
\plotone{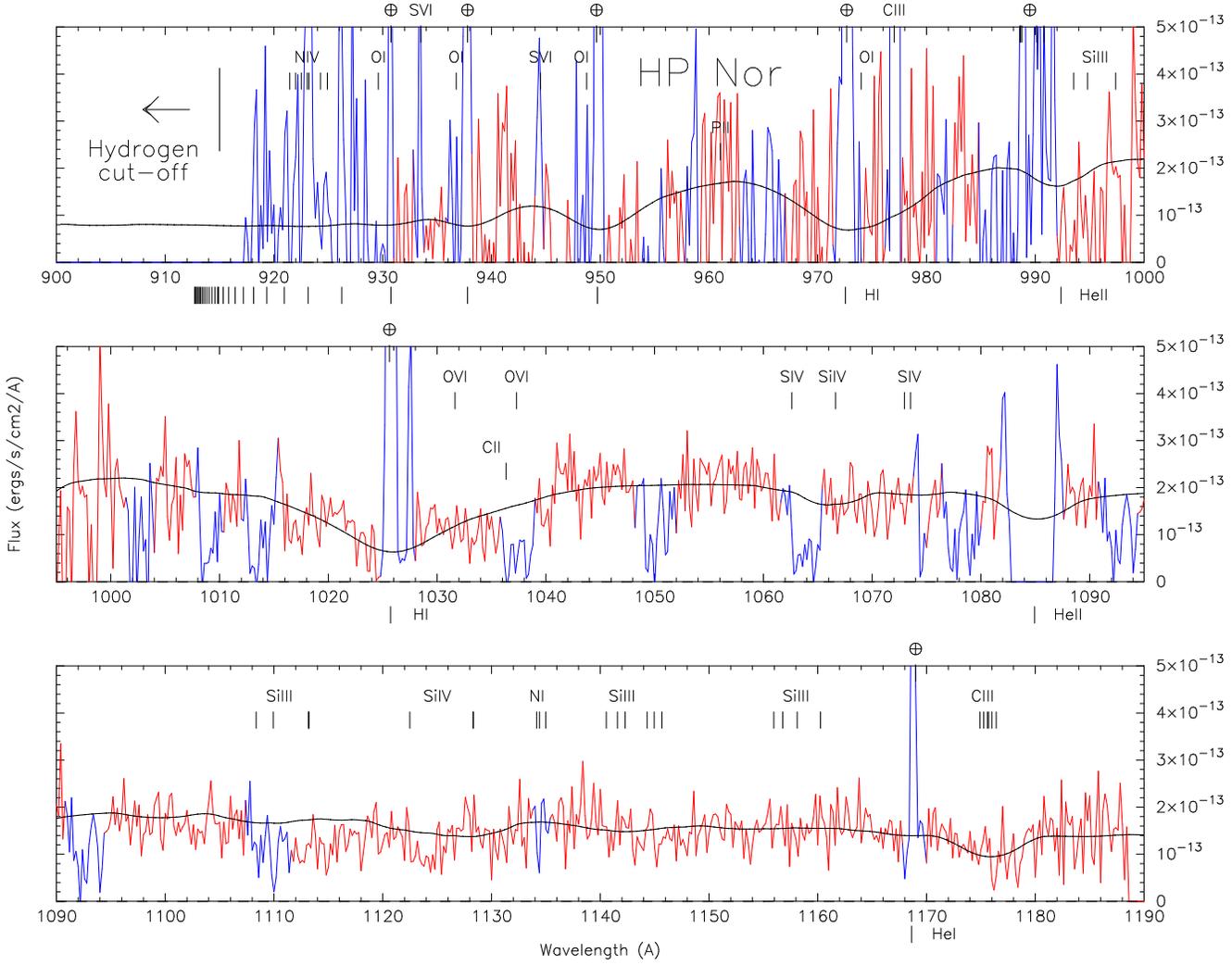} 
\caption{The best fit model to the
{\it FUSE} spectrum of HP Nor assuming E(B-V)=0.2. 
The model is in black, the observed spectrum in red
(light gray), and  
the regions that have been masked before the fitting 
are in blue (dark gray). 
The model consists of a WD with $M=0.8M_{\odot}$, 
T=43,000K, and a projected rotational velocity
$v_{rot} \sin{i}=1000$km/s. 
The distance obtained is d=265pc. 
} 
\end{figure}

\clearpage 

\begin{figure}
\vspace{-5.cm} 
\plotone{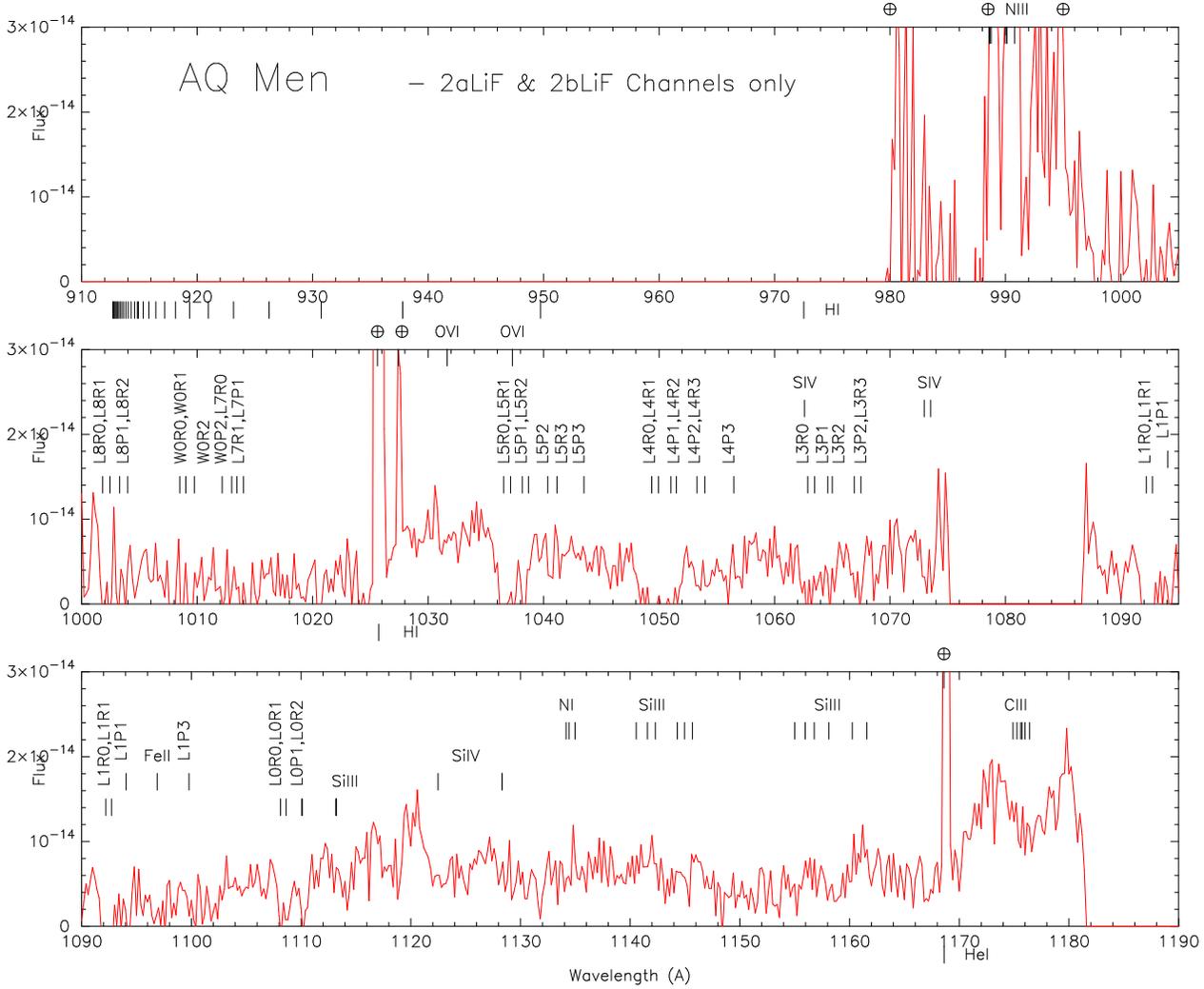}
\caption{The {\it FUSE} spectrum of AQ Men. 
Due to their contamination with air glow and very low signal, the
LiF 1 and SiC 1 \& 2 channels have been discarded. Even though the
signal is very poor, one recognizes the ISM molecular absorption
features (labelled vertically), 
as well as some broad emission from C\,{\sc iii} (1175) and 
the O\,{\sc vi} doublet. One can identify the Si\,{\sc iii} 
(1113\AA) and Si\,{\sc iv} (1123 \& 1128\AA) lines. 
Other metal lines have been marked but are not detected.}  
\end{figure} 

\clearpage 

\clearpage 
\begin{figure}
\vspace{-5.cm} 
\plotone{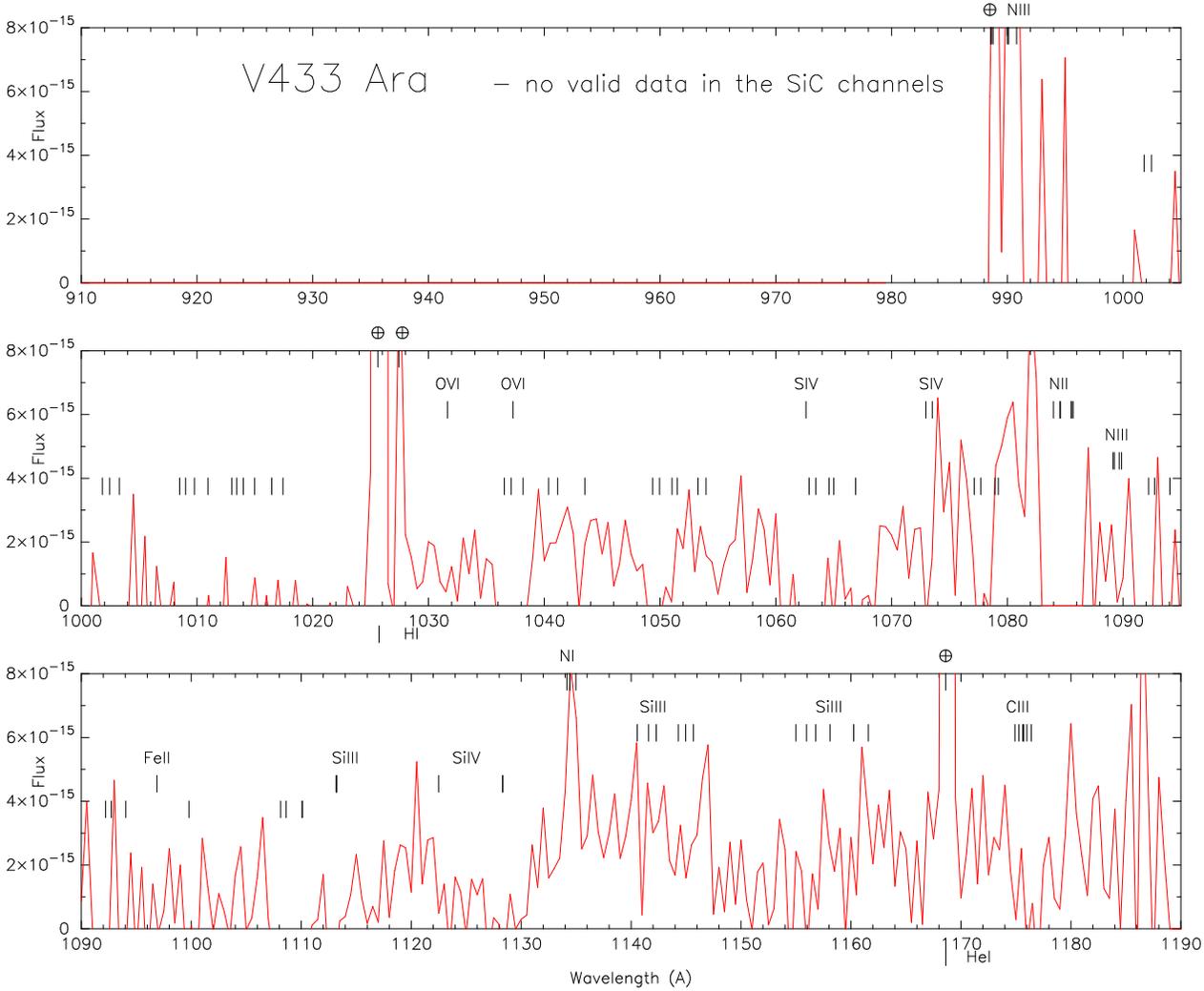}   
\caption{The {\it FUSE} spectrum of V433 Ara.  
Due to their contamination with air glow and very low signal, the
SiC 1 \& 2 channels have been discarded. The S/N is so poor that it
is impossible to model any features, instead the flux level is used to
assess a limit on the  WD temperature and/or the disk mass accretion rate.
We have marked the ISM molecular lines as well as the metals lines
usually seen in spectra of quiescent DNs. Some lines coincide 
(e.g. C\,{\sc iii} 1175\AA\ ) with a dip
in the flux, but no lines are actually detected as the flux also
dips in region where lines are not expected to be observed 
(e.g. 1150\AA).} 
\end{figure}

\clearpage 
\begin{figure}
\vspace{-5.cm} 
\plotone{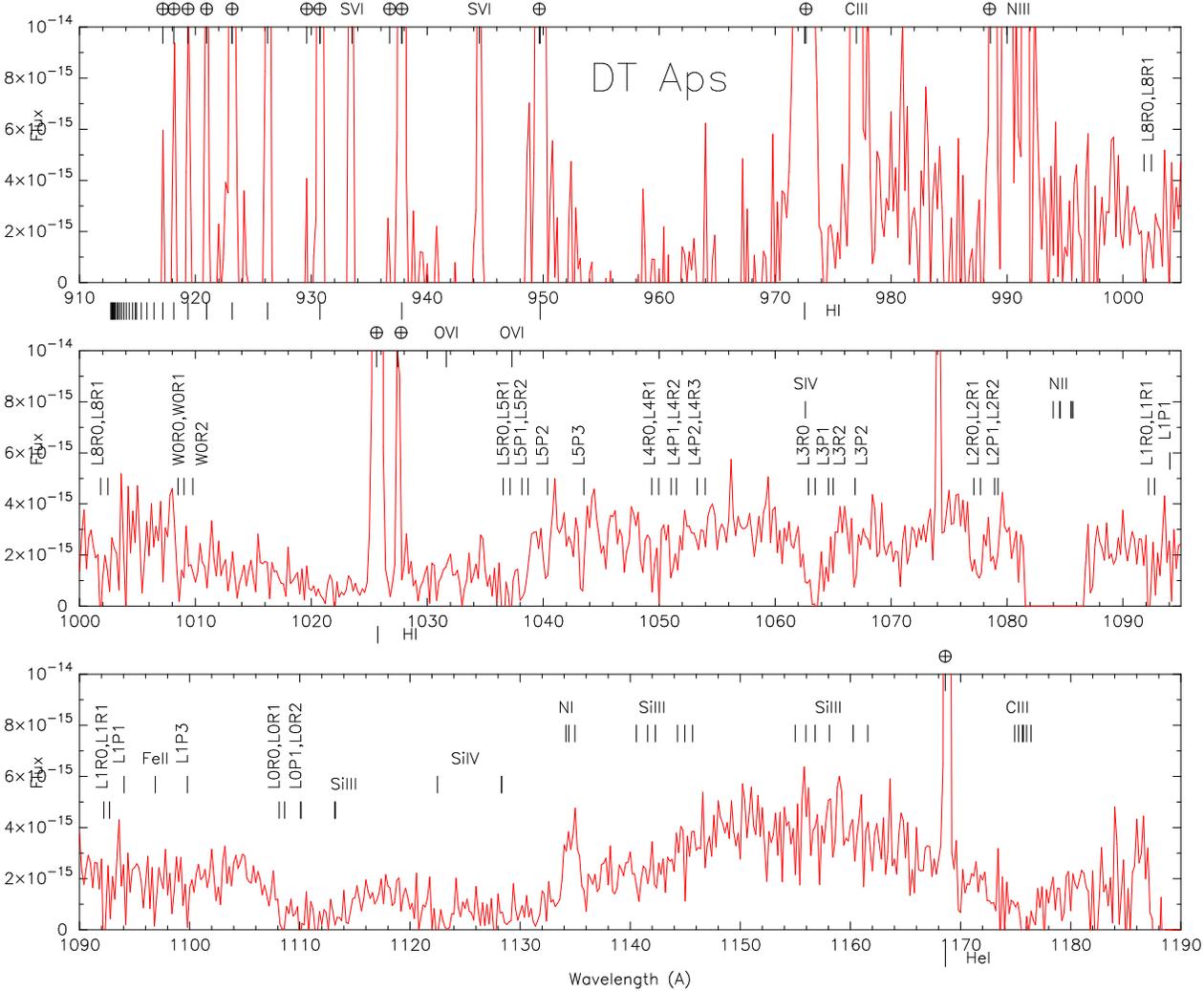}   
\caption{The {\it FUSE} spectrum of DT Aps.
In the upper panel
the spectrum consists mainly of sharp air glow emission lines;
at longer
wavelengths, in the middle panel, the spectrum reveals
a prominent broad and deep Ly$\beta$ absorption feature. 
In the longer wavelengths
(lower panel) the silicon absorption features Si\,{\sc iii} \& Si\,{\sc iv}
(around $\lambda \sim 1110-1130$\AA ) are very deep and
there seems to be some emission around $\lambda \sim 1150-1160$\AA\ ,
both could be due to an incorrect background subtraction.
The broad C\,{\sc iii} ($\lambda 1175$) absorption feature 
is the sign of  a large velocity broadening.
The only lines that are identified are the ISM molecular lines 
and the C\,{\sc iii} (1175\AA) line, all the other lines have been
marked but are not detected. 
Overall the spectrum is rather noisy and of a poor quality. 
} 
\end{figure} 

\clearpage 
\begin{figure}
\vspace{-5.cm} 
\plotone{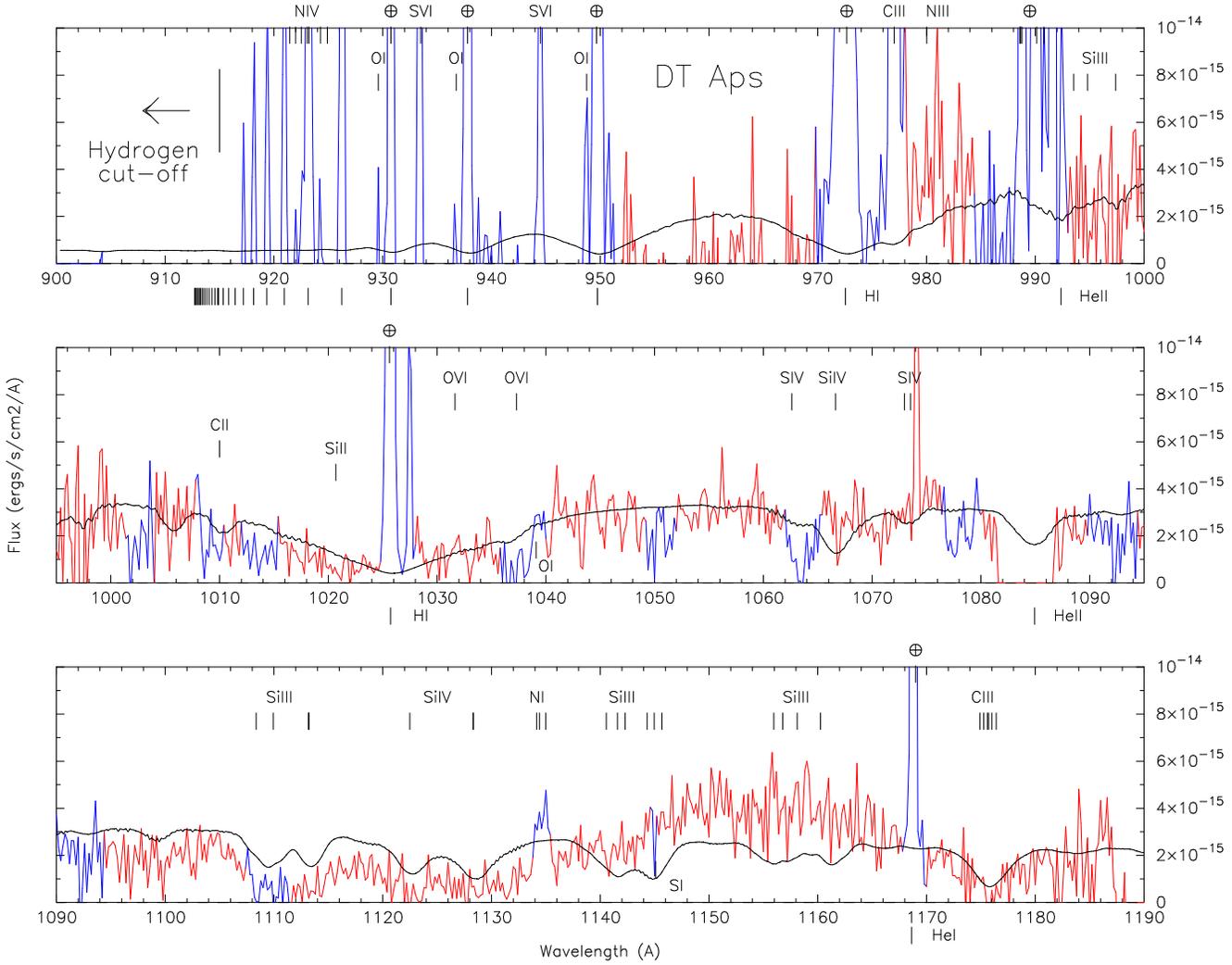}   
\caption{The best fit model (in black) 
to the {\it FUSE} spectrum (in red/light grey)
of DT Aps. The regions that have been masked before fitting are
in blue/dark grey (they consist mainly of ISM molecular hydrogen absorption
lines and air glow emission). . 
The model consists of a WD with a mass $M=0.8M_{\odot}$, 
with T=37,000K, a silicon abundance of $10 \times$ solar,  
d=1700pc and $\chi^2 = 0.2050$.
In order to fit the absorption lines profiles 
the projected rotational velocity $V_{rot} \sin{i}$ was set to 500km/s.
Here we assumed $E(B-V)=0.0$.
} 
\end{figure} 

\clearpage 

\begin{figure}
\vspace{-5.cm} 
\plotone{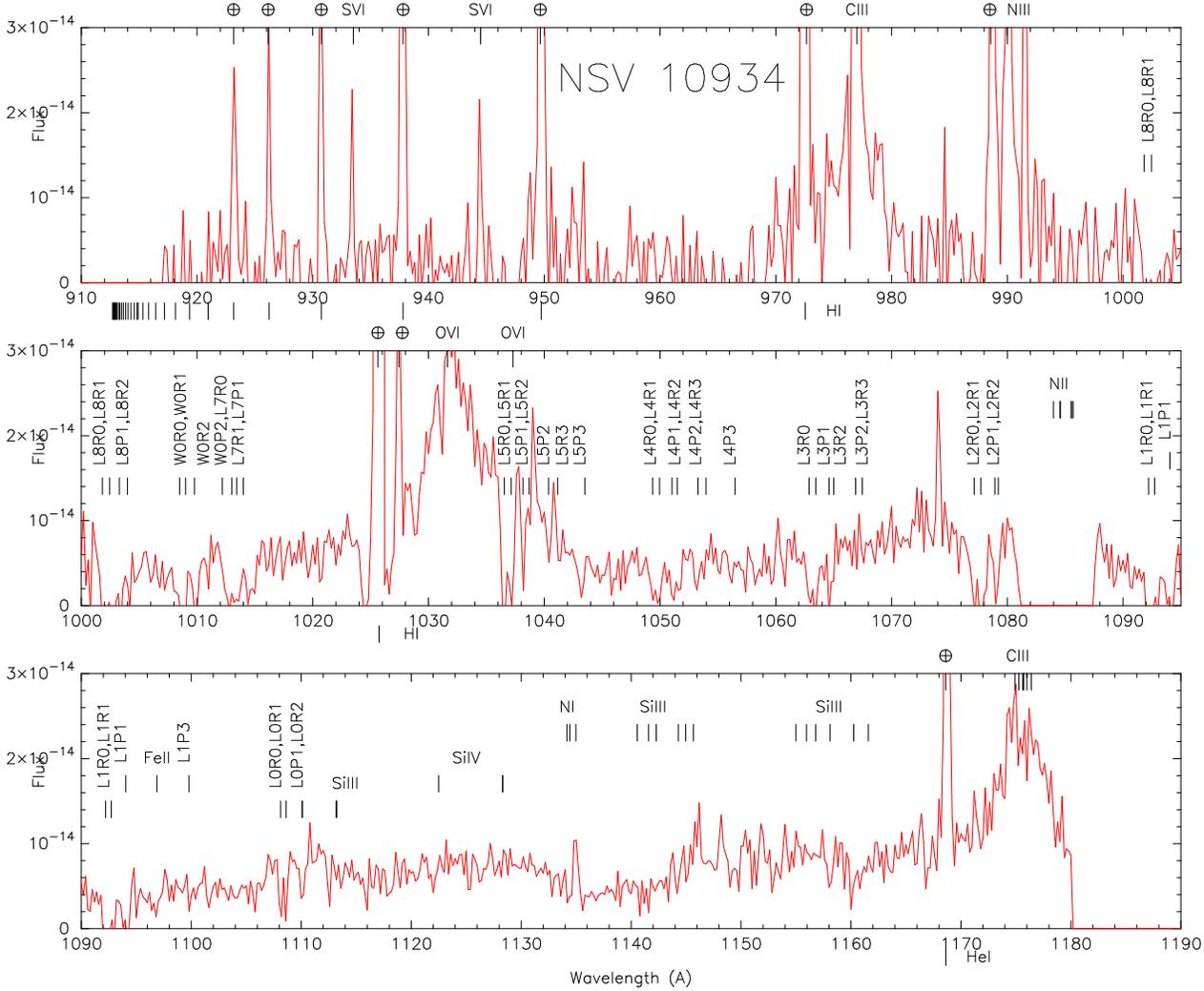}   
\caption{The {\it FUSE} spectrum of NSV 10934 has a low 
S/N especially in the shorter wavelengths 
due to the fact that the data from the
2bSiC, 2bLiF and 1aSiC channels were unusable and a large portion
of the 1bLiF channel was lost to the worm. As a consequence there
is gap around 1085\AA .  
The continuum is rather flat with very broad emission lines
from C\,{\sc iii} ($\lambda$977 \& $\lambda$1175) and the O\,{\sc vi}
doublet.
The Ly$\beta$ broad absorption feature is not detected, possibly
due to the left wing of the very broad oxygen emission feature
or to velocity broadening in a disk viewed at a high inclination.
The ISM molecular lines are identified (labelled vertically).
The metal lines have been marked but are not detected. 
All the sharp emission lines are due to air glow. 
} 
\end{figure}

\clearpage 

\begin{figure}
\vspace{-5.cm} 
\plotone{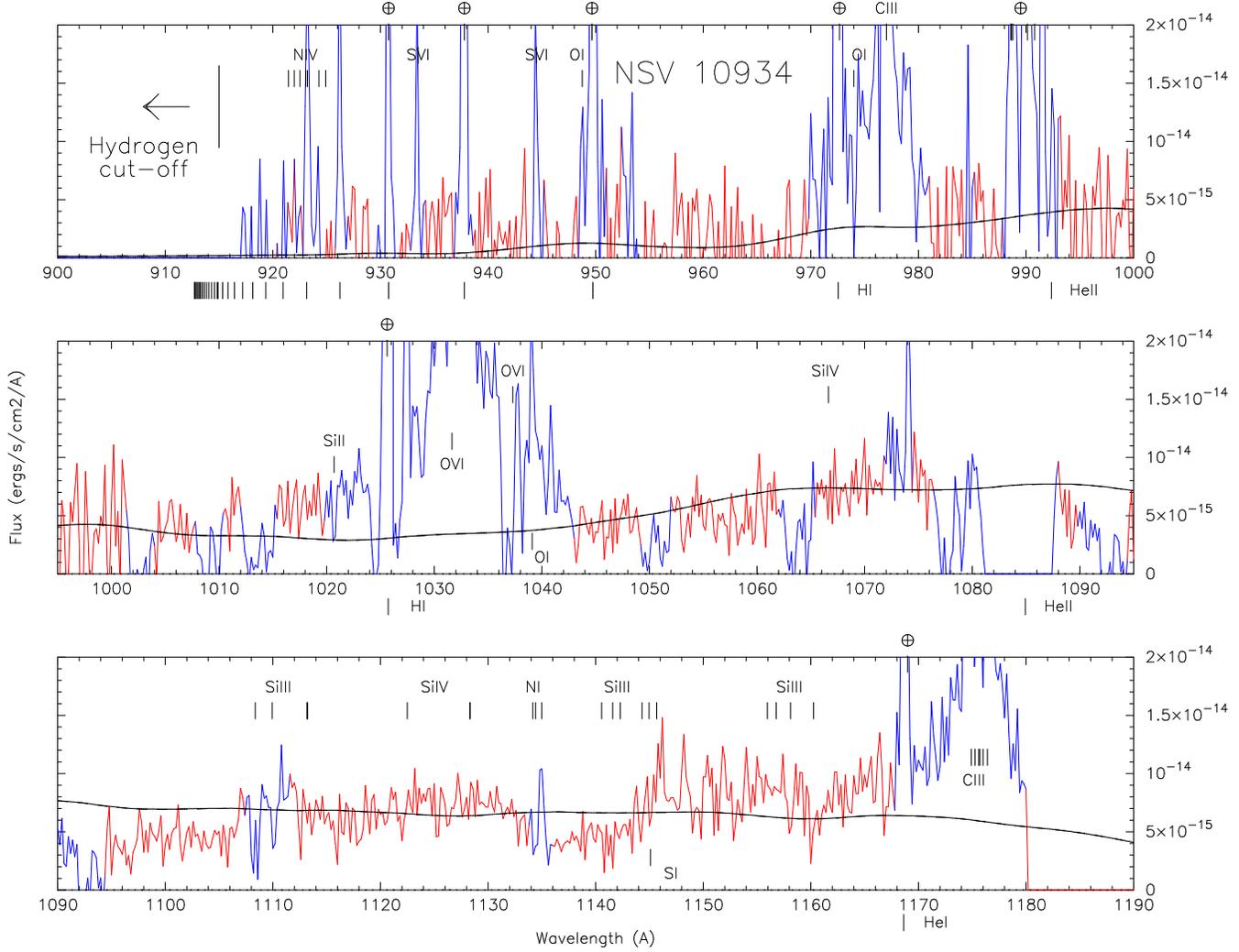}   
\caption{The best fit model to the {\it FUSE} spectrum of NSV 10934.
The observed spectrum (in red/light gray) 
has been dereddened assuming E(B-V)=0.1.
The regions that have been masked before modeling are shown in
blue (dark gray).  
The best fit (in black) consists of a disk model with a 
$M=1.2M_{wd}$, $\dot{M}=10^{-10}$, and $i=81^{\circ}$, giving  a
distance of 187pc $\chi^2=0.146$.   
} 
\end{figure}

\clearpage 
\begin{figure}
\vspace{-5.cm} 
\plotone{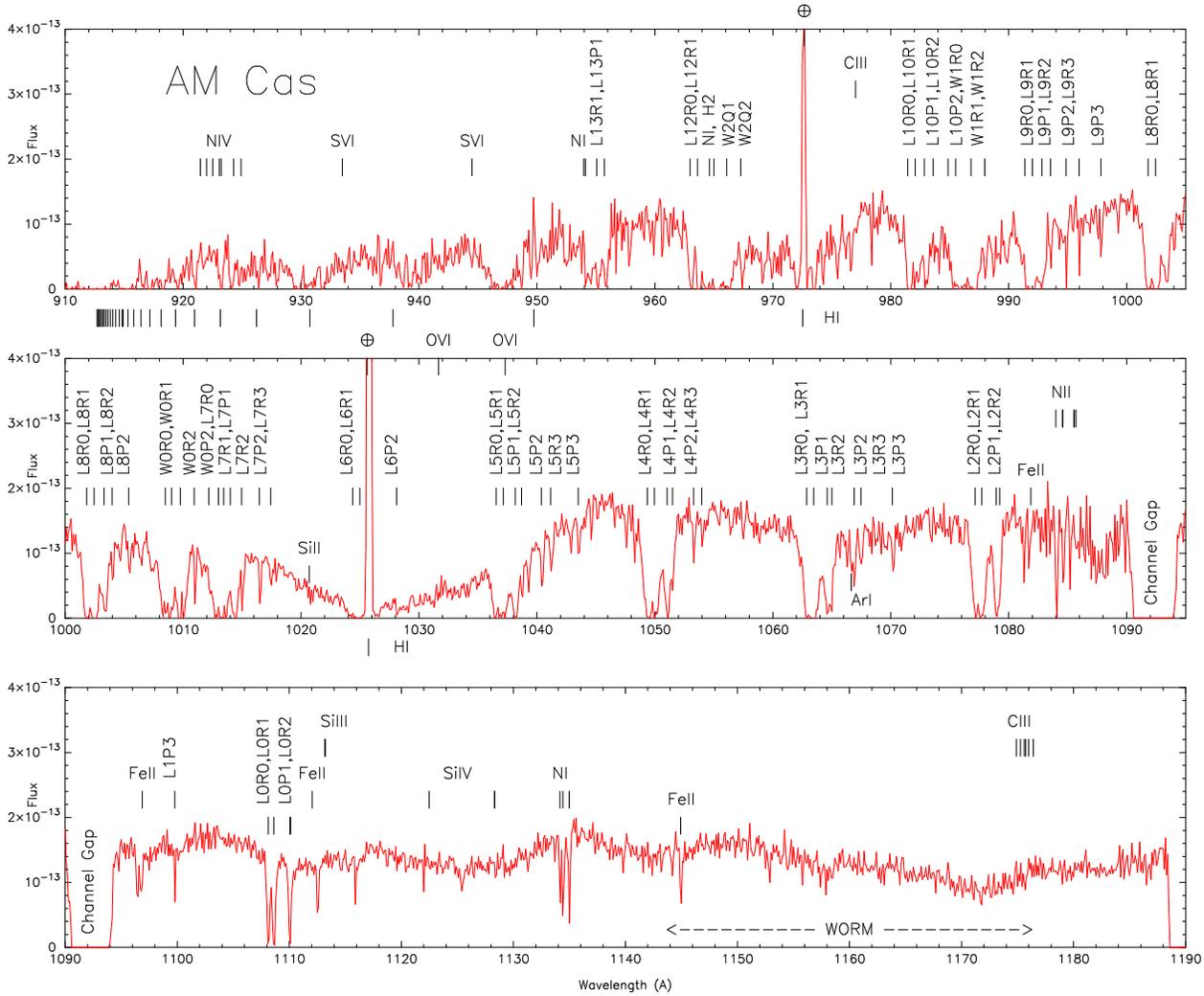}   
\caption{
The {\it FUSE} spectrum of AM Cas with line identifications.
The ISM molecular hydrogen absorption lines have been labeled
vertically.  
This spectrum consists only of the 1-SiC and -LiF a \& b channels,
because of that there is a gap around 1095\AA , and {\bf the longer
wavelengths are affected by the worm} (as shown). 
The higher ionization level
metal lines are discussed in Fig.11 in the context of the fit
model.    
} 
\end{figure}

\clearpage 
\begin{figure}
\vspace{-5.cm}
\plotone{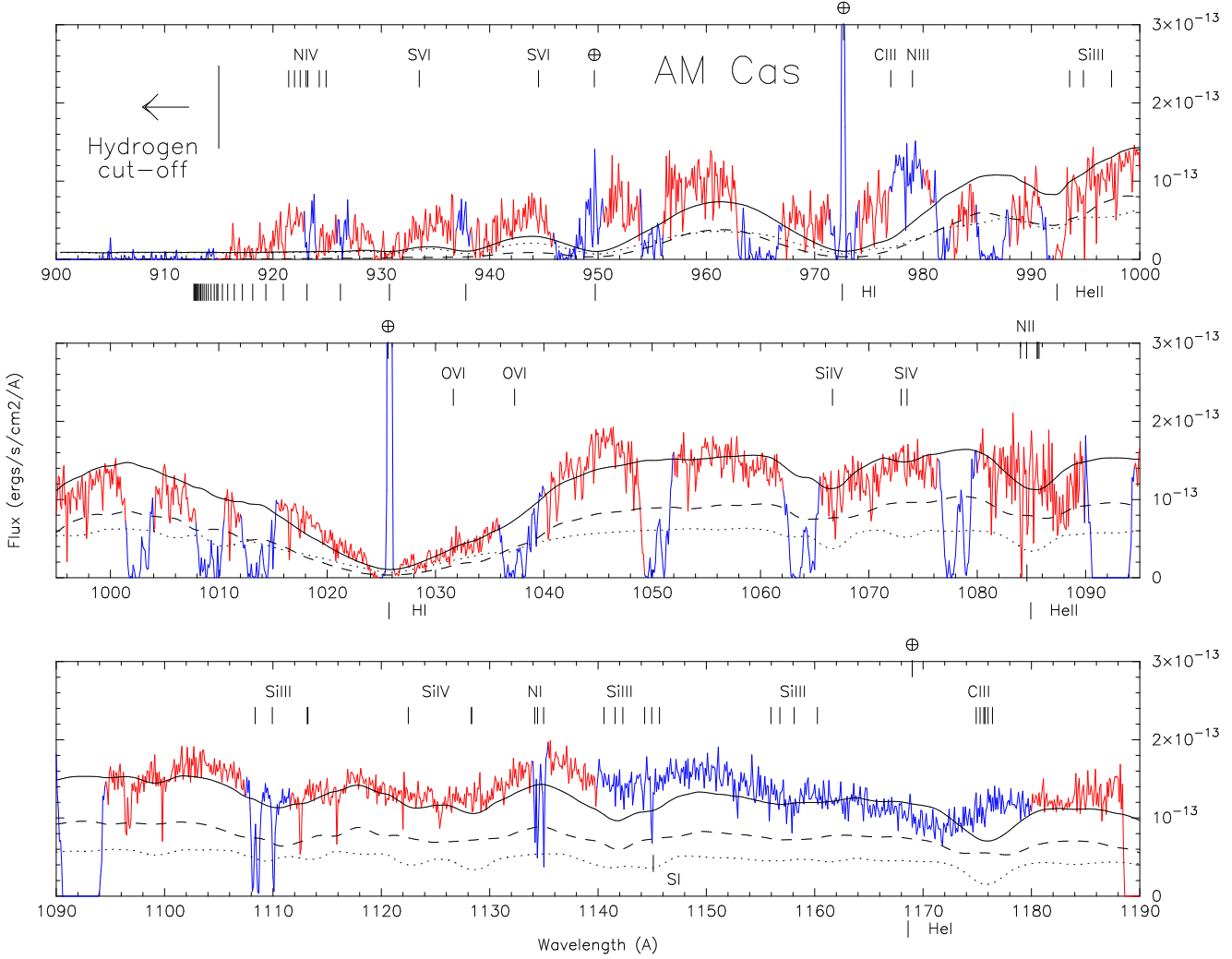}   
\caption{The best fit model (solid black line) 
to the {\it FUSE} spectrum of AM Cas (in light grey/red) assuming
E(B-V)=0.0 is a WD+disk composite model. 
The regions that have been masked before fitting are shown
in dark grey (blue). 
The WD model (dotted line) has $M=0.8M_{\odot}$, $T=36,000$K,  
$V_{rot} \sin{i}$=500km/s, and the disk model (dashed-line) has a mass accretion
rate $\dot{M}= 2 \times 10^{-10}M_{\odot}$/yr and  
an inclination $i=18^{\circ}$. This model gave a distance of
373pc,  $\chi^2=0.563$, with the WD contributing 41\% of the
flux and the disk 59\%.}  
\end{figure} 

\clearpage 
\begin{figure}
\vspace{-5.cm}
\plotone{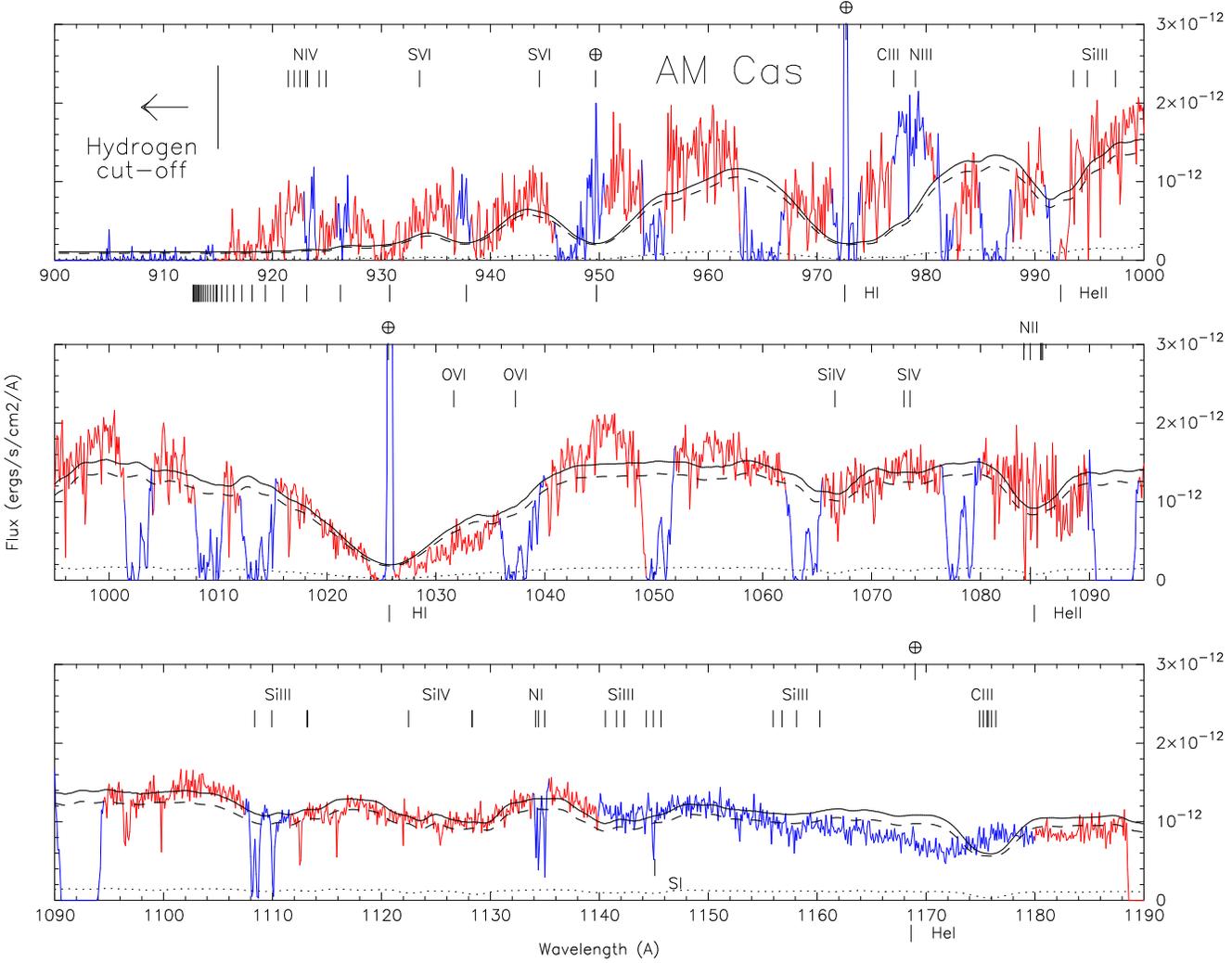}       
\caption{The best fit model (solid black) to the {\it FUSE} spectrum of AM Cas assuming
E(B-V)=0.2 is a WD+disk composite model. 
The WD model (dotted line) has $M=0.55M_{\odot}$, $T=35,000$K,  
$V_{rot} \sin{i}$=400km/s, and the disk model (dashed line) 
has a mass accretion
rate $\dot{M} = 3 \times 10^{-9}M_{\odot}$/yr and  
an inclination $i=18^{\circ}$. This model gave a distance of
331pc,  $\chi^2=0.538$, with the WD contributing 11\% of the
flux and the disk 89\%. The excess flux in the shorter wavelengths
could be due to emission from N\,{\sc iv} and S\,{\sc vi}.
There seems to be some broad C\,{\sc iii} (977\AA) \& 
N\,{\sc iii} (980\AA) emission, while the higher flux
around  $\sim$950-962\AA\ cannot be accounted for. 
In the lower panel the Si\,{\sc iii} (1108-1113\AA), 
Si\,{\sc iv} (1123 \& 1128\AA), and Si\,{\sc iii} (1140-1145\AA)
lines are broadened (Keplerian motion) and are identified as the
depression in the flux there. The Si\,{\sc iii} ($\sim$1158\AA) 
and C\,{\sc iii} (1175\AA) are not identified, possibly due to
the deterioration of the spectra in the region of the worm.  
}  
\end{figure}

\clearpage 
\begin{figure}
\vspace{-5.cm} 
\plotone{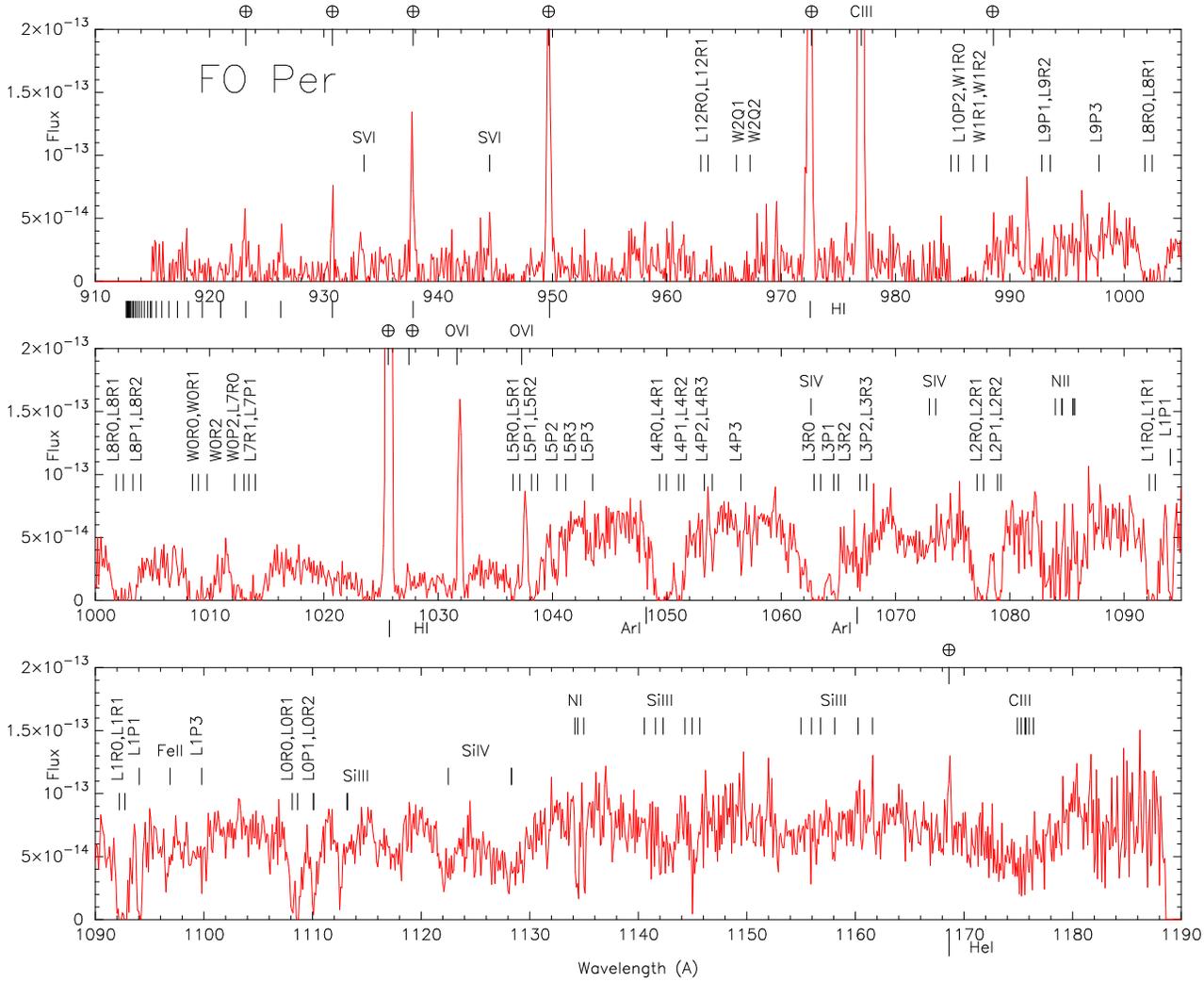}       
\caption{The {\it FUSE} spectrum of FO Per. All the sharp
emission lines are due to geo- and helio-coronal contamination,
including the S\,{\sc vi} and O\,{\sc vi} lines. All the ISM hydrogen
molecular absorption lines have been annotated vertically.
The broad Ly$\beta$ feature is visible, as well as
broad carbon (C\,{\sc iii} 1175\AA) 
and silicon (Si\,{\sc iii} 1113\AA, Si\,{\sc iv} 1123 \& 1128\AA, 
Si\,{\sc iii} $\sim$1140-1145\AA) absorption features.
All the other metal lines have not been identified, but have been
marked for comparison. 
}  
\end{figure}

\clearpage 
\begin{figure}
\vspace{-5.cm} 
\plotone{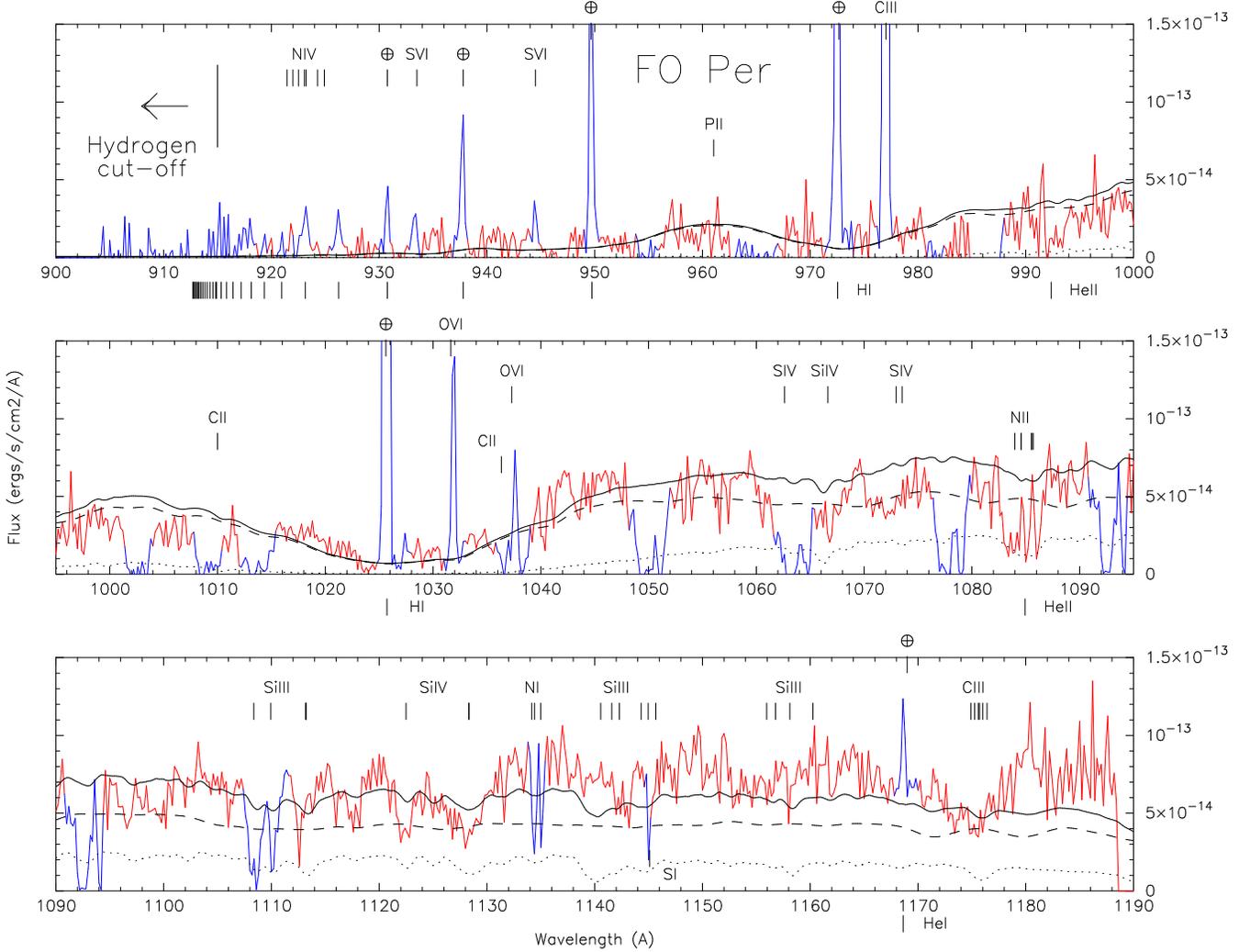}       
\caption{The best-fit WD+disk composite model 
(solid black) to the 
{\it FUSE} spectrum of FO Per assuming E(B-V)=0.0.
The observed spectrum is in red (light gray) and
the regions that have been masked for the fitting
are in blue (dark gray). 
The WD model (dotted line) has  $0.4M_{\odot}$, with
$T_{wd}=21,000$K, and $V_{rot} \sin{i}=$200km/s,
the disk model (dashed line) has
$i=75^{\circ}$, and  $\dot{M} = 10^{-8.5}M_{\odot}$yr$^{-1}$.
The distance obtained is d=291pc, and  $\chi^2=0.258$.
The WD contributes 29\% of the FUV flux while the disk
contributes 71\%.  } 
\end{figure}

\clearpage 
\begin{figure}
\vspace{-5.cm} 
\plotone{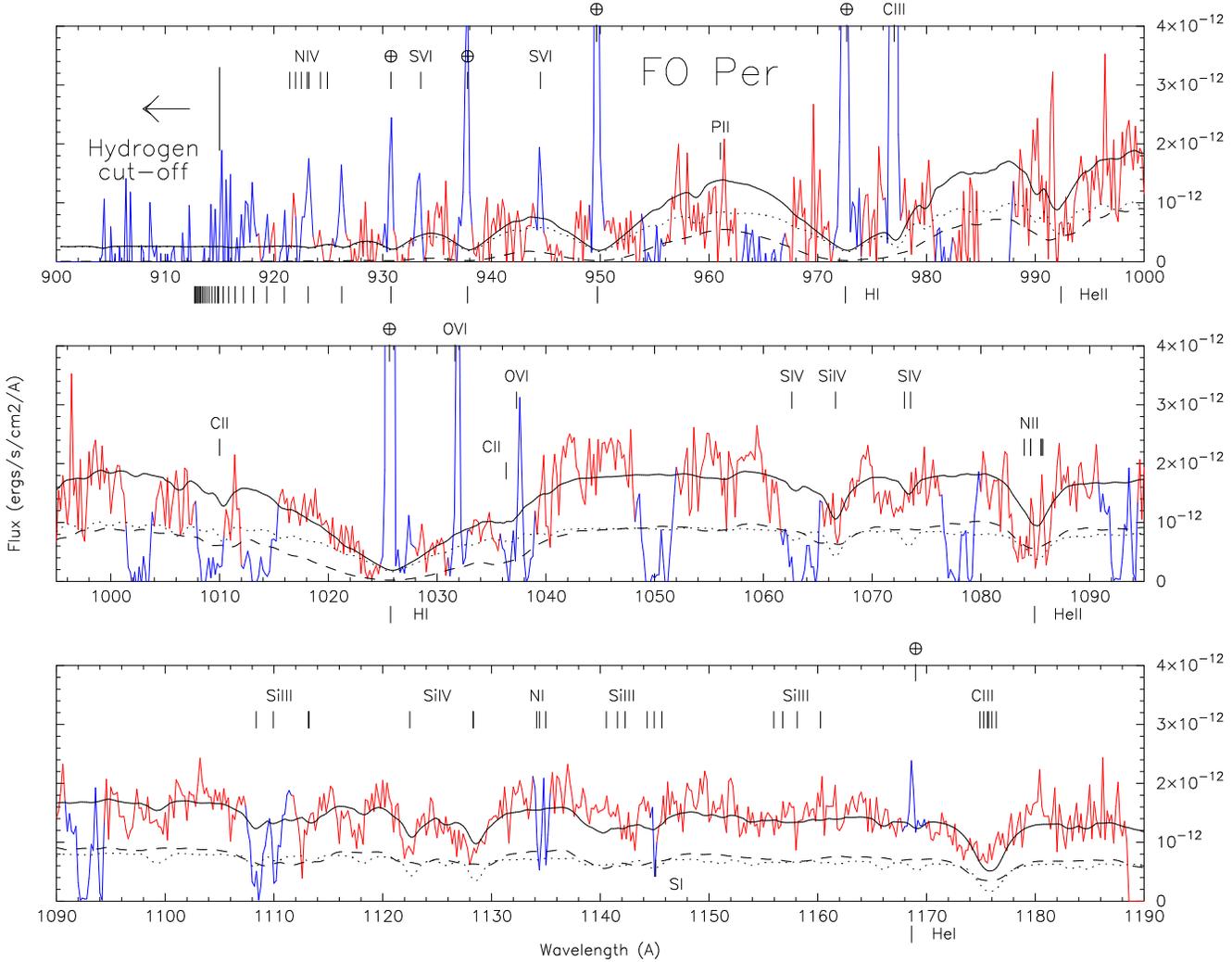}       
\caption{The best-fit WD+disk composite model to the 
dereddened {\it FUSE} spectrum of FO Per assuming E(B-V)=0.3.
The WD model (dotted line) has a $0.4M_{\odot}$ and  
$T_{wd}=40,000$K, and $V_{rot} \sin{i}=$200km/s.
The disk model (dashed line) has 
$i=18^{\circ}$, and $\dot{M} = 10^{-8.5}M_{\odot}$yr$^{-1}$.
The model gives a distance d=254pc with $\chi^2=0.146$.
The WD contributes 52\% of the FUV flux while the disk
contributes 48\%. When compared to the model, we find that the
C\,{\sc iii} (1175\AA) and Si\,{\sc iv} (1123, 1128\AA) lines are
blue shifted by $\sim$2\AA\ and $\sim$1\AA\ respectively.
The S\,{\sc iv} (1063,1073\AA) lines might also be blue-shifted.
} 
\end{figure}

\clearpage 
\begin{figure}
\vspace{-5.cm} 
\plotone{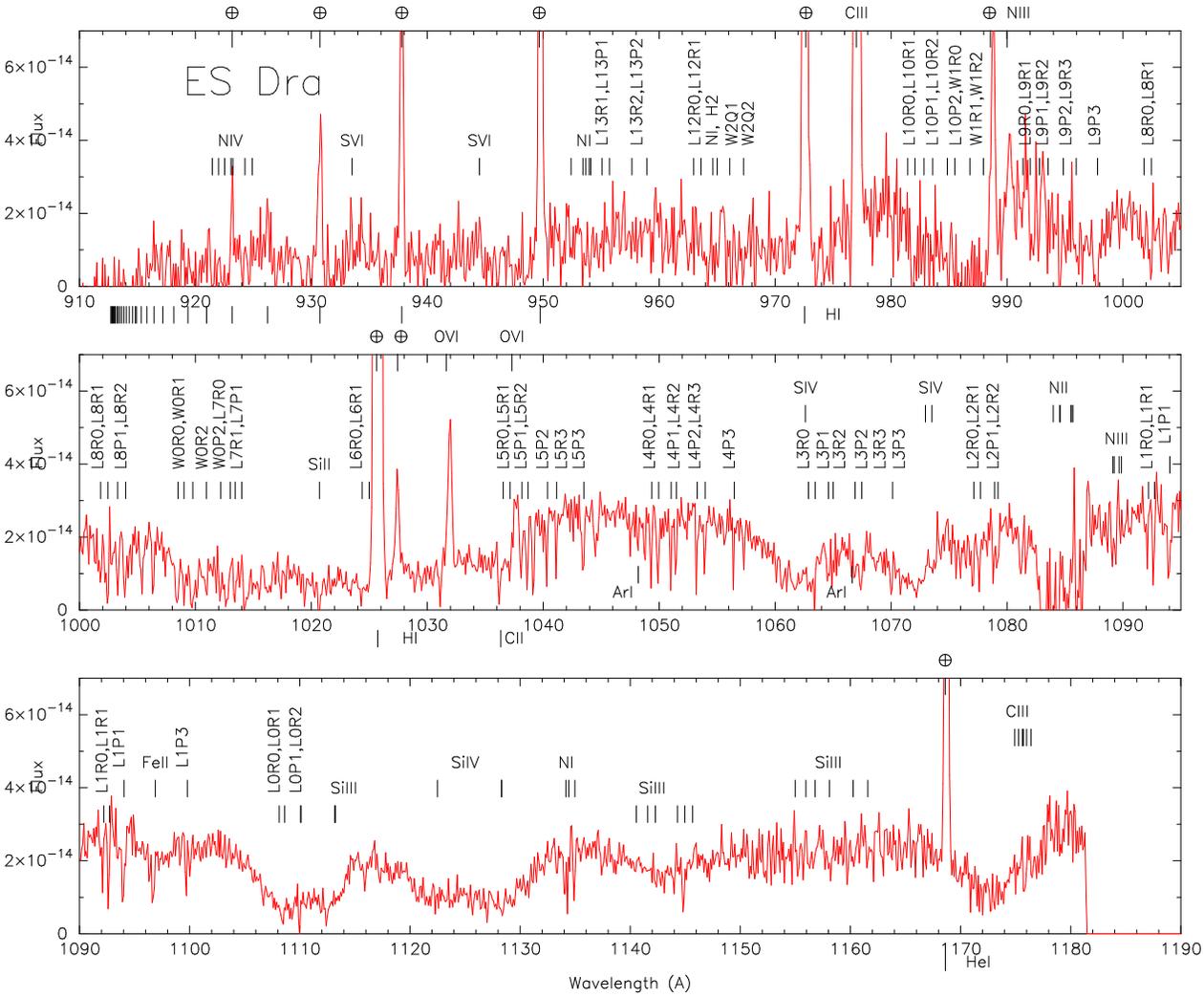}       
\caption{The {\it FUSE} spectrum of ES Dra with line
identifications. All the sharp emission lines are due to 
geo- and helio-coronal contamination, including the 
S\,{\sc vi}, O\,{\sc vi}, and C\,{\sc iii} (977\AA) lines.  
Broad absorption features from carbon and
silicon are clearly seen, slightly blue-shifted. The shitfing of the
lines is more apparent in Figs.16 \& 17, where the observed spectrum
is compared to the models.} 
\end{figure}

\clearpage 
\begin{figure}
\vspace{-5.cm} 
\plotone{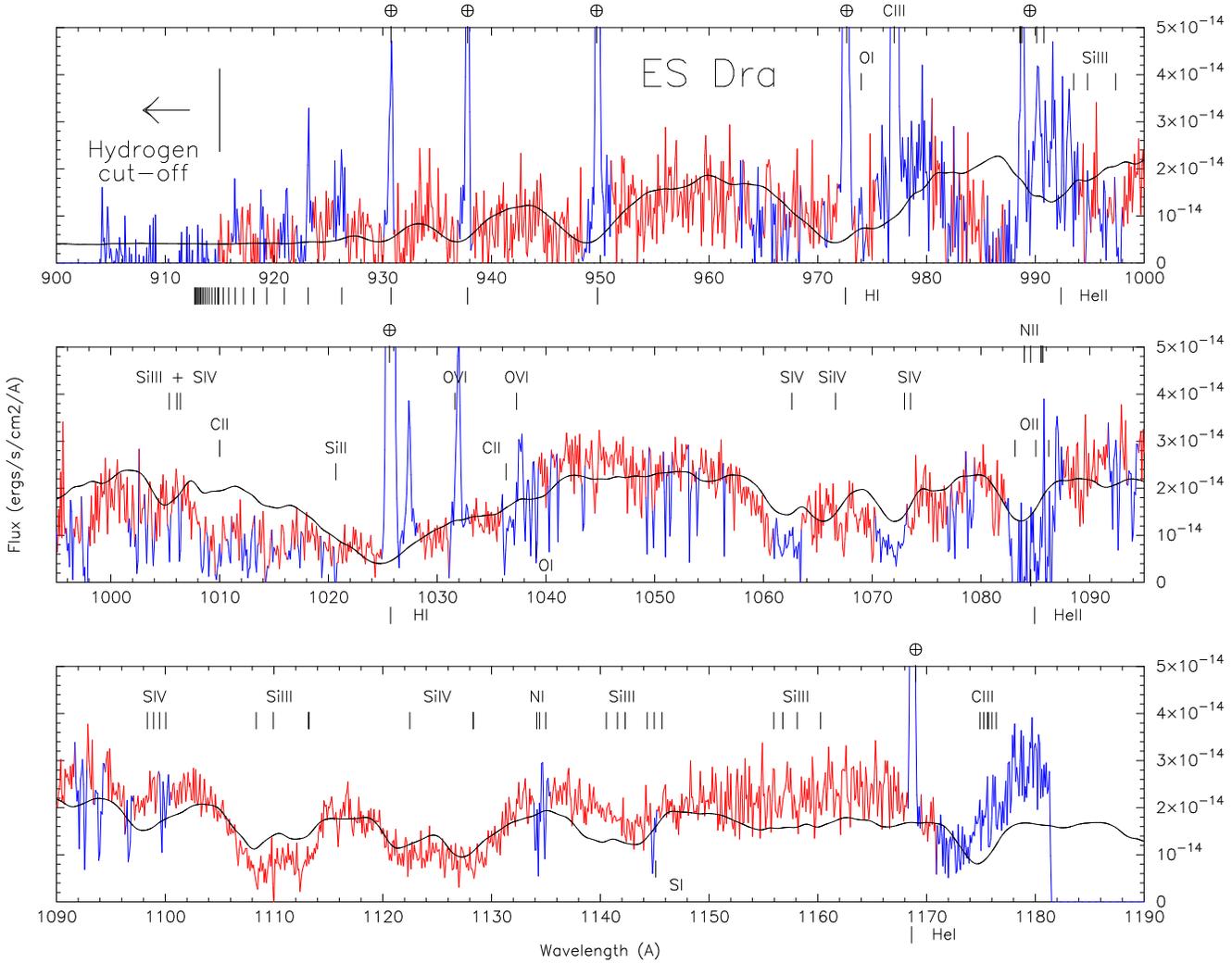}       
\caption{The best-fit accretion disk model to the {\it FUSE}
spectrum of ES Dra. The regions of the spectrum that have been
masked for the fitting are in blue.  
The disk model has a central accreting WD with 
a mass $M=1.2M_{\odot}$, 
a mass accretion rate $\dot{M} = 10^{-9.5}M_{\odot}$yr$^{-1}$, 
an inclination $i=5^{\circ}$,
and gives a distance of 1754pc.  
We fine-tuned the fitting by setting the silicon and sulfur 
abundances $ 10 \times$ solar and $ 50 \times$ solar respectively,
and we shifted the entire synthetic spectrum 
to the blue by 1.0\AA . This resulted in  $\chi^2=0.4280$.}
\end{figure}

\clearpage 
\begin{figure}[thp] 
\vspace{-5.cm} 
\plotone{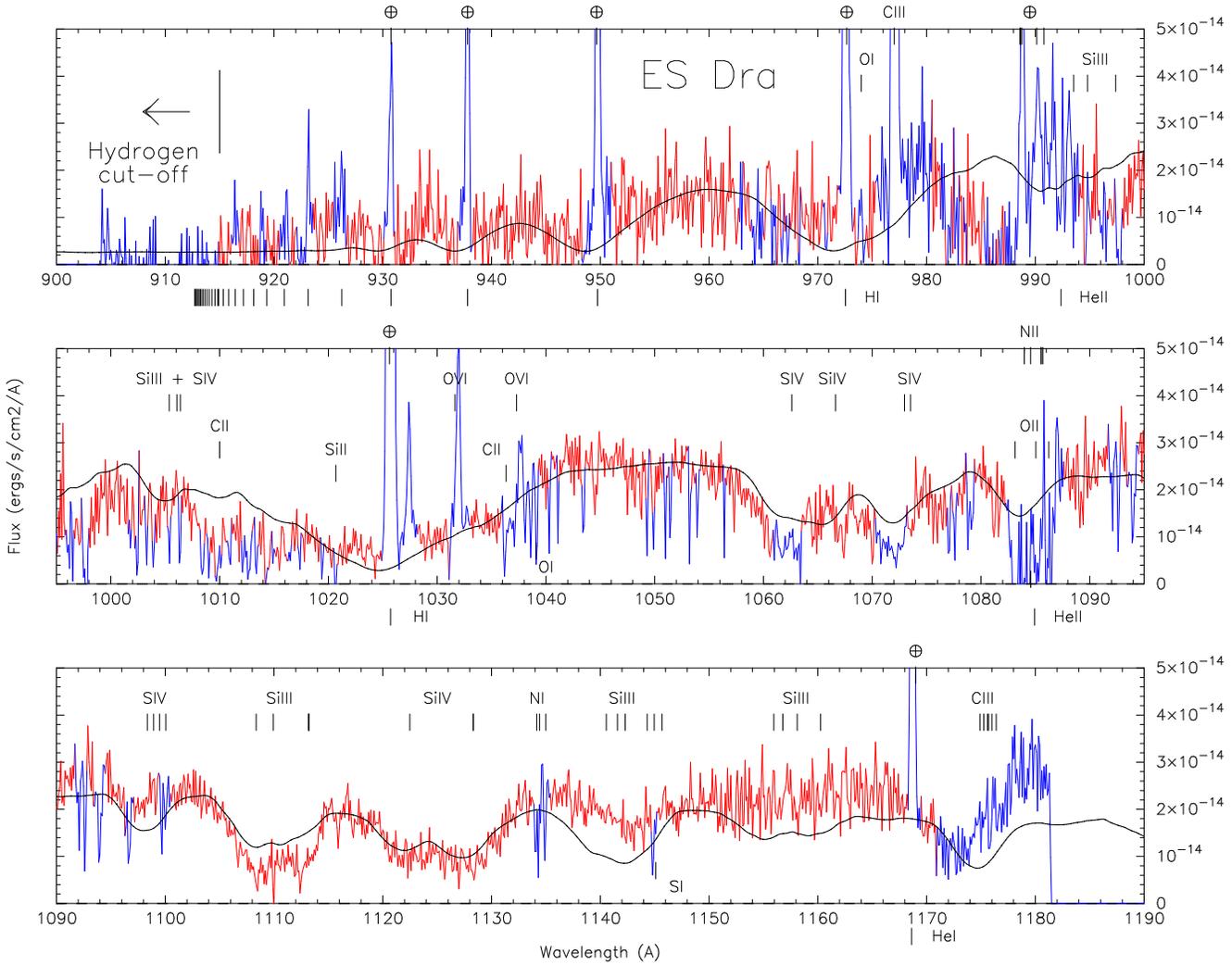}    
\caption{The best fit single component WD model to the {\it FUSE} spectrum
of ES Dra. The WD has a temperature 
$T=35,000$K,
$log(g)=7.8$ (corresponding to $M_{wd} \approx 0.58M_{\odot}$),
a projected rotational velocity $v_{rot} \sin{i} = 700$km/s, 
abundances $Si \approx 10 \times$ solar, $S \approx 50 \times$ solar and a
blue-shift of 1.3\AA .
This model gives a distance of 773pc and $\chi^2 = 0.4277$.
When compared to the synthetic spectrum, the observed spectrum 
clearly reveals the Si\,{\sc iii} ($\sim$1110\AA) and 
Si\,{\sc iv} (1023 \& 1028\AA) lines; the Si\,{\sc iii} lines
in the longer wavelenghts do not match the model
The C\,{\sc iii} (1175\AA) presents a P-Cygni profile; the 
S\,{\sc iv} (1063, 1073, \& 1099\AA)  lines are blue-shifted. 
The Si\,{\sc iii} + S\,{\sc iv} ($\sim$1006\AA), 
C\,{\sc ii} (1010\AA), Si\,{\sc iv} (1066\AA) lines are all contaminated
with ISM absorption (see Fig.15 for comparison).
}
\label{fig:esdrawd} 
\end{figure} 

\clearpage 
\begin{figure}
\vspace{-9.cm} 
\plotone{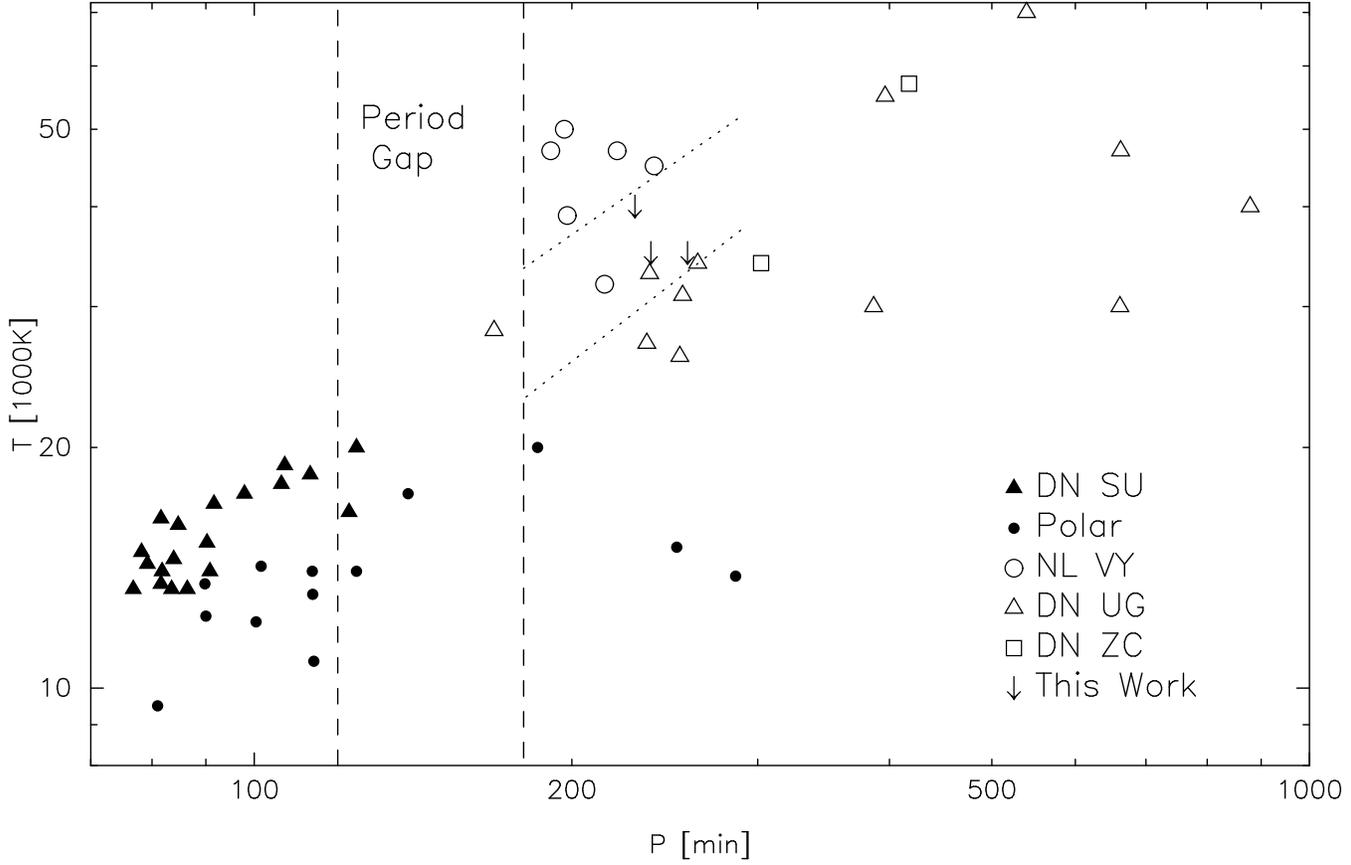} 
\caption{Effective White Dwarf Temperature as a function of the orbital
period, from Table 5. 
There is a well defined separation between Polars and DN below the gap,
and apparentlly also above the gap.
There is clearly a lack of data points above the
gap for Z Cam's, Polars and VY Scl's.  
There seems to be a 
separation in the $P-T_{eff}$ parameter space between Polars, 
SU UMa's, U Gem's and possibly VY Scl's.  
The 3 arrows (this work) are only upper limits and
have been included here to show how important it is to obtain good FUV
spectra and parallax for CVs in order to be able to derive robust
results for the WD effective temperatures. The traditional magnetic
braking above the period gap \citep{how01} is shown between the dotted
lines.   
} 
\end{figure}

\end{document}